\documentclass[%
 amsmath,amssymb,
 aps, 
 reprint,
prd,
onecolumn,
superscriptaddress,
floatfix,
raggedbottom, 
nofootinbib,
longbibliography, 
 notitlepage,
]{revtex4-1}
\usepackage[T1]{fontenc}
\usepackage[utf8]{inputenc}
\usepackage{xcolor}
\usepackage{bm}
\usepackage{amsmath}
\usepackage{mathtools}
\usepackage{amssymb}
\usepackage[english]{babel}
\usepackage{float}
\usepackage[margin=1.3in]{geometry}

\usepackage[%
colorlinks=true,    
citecolor=blue,     
linkcolor=black,    
urlcolor=black,      
]{hyperref}

\makeatletter
\newread\bibcache@input
\newwrite\bibcache@output
\newif\ifbibcache@stale
\newcommand{\bibcache@discardif}[2]{%
  \IfFileExists{#1}{%
    \begingroup
    \bibcache@stalefalse
    \edef\bibcache@needle{\detokenize{#2}}%
    \openin\bibcache@input=#1\relax
    \loop\unless\ifeof\bibcache@input
      \readline\bibcache@input to\bibcache@line
      \edef\bibcache@test{\noexpand\in@{\bibcache@needle}{\bibcache@line}}%
      \bibcache@test
      \ifin@\bibcache@staletrue\fi
    \repeat
    \closein\bibcache@input
    \ifbibcache@stale
      \immediate\openout\bibcache@output=#1\relax
      \immediate\closeout\bibcache@output
      \typeout{Discarded incompatible biblatex cache: #1}%
    \fi
    \endgroup
  }{}%
}
\bibcache@discardif{\jobname.aux}{abx@aux@}
\bibcache@discardif{\jobname.bbl}{biblatex auxiliary file}
\makeatother

\begin{document}

\title{\textbf{Generalization properties of the token perceptron: the role of embedding and vocabulary}}

\author{Dario Bocchi}
\email{dario.bocchi@uniroma1.it}
\affiliation{Physics Department, University of Rome Sapienza, Rome, Italy}
\affiliation{Institute of Nanotechnology, CNR-NANOTEC, Rome, Italy}

\author{Matteo Negri}
\email{matteo.negri1@cyu.fr}
\affiliation{Laboratoire de Physique Théorique et Modélisation, CY Cergy Paris Université}

\author{Flavio Nicoletti}
\email{flavion@chalmers.se}
\affiliation{Department of Computer Science and Engineering Chalmers University of Technology and University of Gothenburg, Sweden}

\author{Francesco D'Amico}
\email{francesco.damico@uniroma1.it}
\affiliation{Physics Department, University of Rome Sapienza, Rome, Italy}
\affiliation{Institute of Nanotechnology, CNR-NANOTEC, Rome, Italy}

\begin{abstract}
\hspace{-10pt}Large language models are trained to predict tokens from a finite vocabulary, each represented by a vector in a $d$-dimensional embedding space. Therefore, token prediction is a classification problem whose classes have a geometry determined by the embedding. We study how this geometry affects learning in a minimal solvable model: a linear predictor that maps a sequence of $N$ embedded tokens to a $d$-dimensional output through one $d\times d$ matrix per position, trained in the teacher--student setting. We call this the \textit{token perceptron}. We compute its generalization error with the replica method, taking $N\to\infty$ at fixed number of examples per position $\alpha=P/N$ and keeping $d$ and the vocabulary size $D$ finite. As a first reference case, we consider continuous embeddings on the sphere, which isolate the role of the embedding dimension. The natural load is then $\alpha/d$, and for $d>1$ a normalized label fixes the direction of the teacher output, so the problem becomes a noiseless regression: the teacher is recovered exactly above $\alpha_{\rm rec}=d^2/(d-1)$, and spherical cross-entropy reaches this threshold at large inverse temperature. In the same setting the storage capacity of random associations scales as $d^2$ for margins $\mathcal{O}(1/\sqrt{d})$. We then turn to vocabulary embeddings, where a label only identifies the Voronoi cell of the teacher output, so that each example imposes inequalities rather than equations, as in the classical perceptron. We compute the Bayes-optimal error for a given embedding matrix through an approximate spectral closure and show that the optimal learner first resolves the directions in which the vocabulary has the largest variance, even when the teacher is isotropic.
\end{abstract}

\maketitle

\setlength{\parindent}{-0.05em}
\setlength{\parskip}{10pt}

\section{Introduction and related work}

Large language models (LLMs) are trained, before any fine-tuning or reinforcement-learning stage~\cite{NEURIPS2022_b1efde53}, to predict the next token of a text \cite{radford2018improving, NEURIPS2020_1457c0d6}. Each token belongs to a finite vocabulary and is represented by a vector in a $d$-dimensional embedding space. The network maps a sequence of such vectors to an output vector, which is decoded into a token by comparing it with the learned vocabulary embeddings; in models with tied input and output embeddings the same vectors serve both roles \cite{NIPS2017_3f5ee243, press2017output}. Token prediction is therefore a classification problem whose classes have a geometry determined by the embedding. In this work we study how this geometry, namely the embedding dimension $d$, the vocabulary size $D$ and the arrangement of the vocabulary vectors, produces an effect on the learning process, as for example the amount of data needed to learn the task.

For the perceptron, questions of this kind received exact answers from the statistical mechanics of learning developed between the late 1980s and the late 1990s: the number of random associations a network can store \cite{E_Gardner_1988, gardner1988optimalstorage}, the generalization error as a function of the number of examples in the teacher--student setting \cite{sompo_perc, Engel2001StatisticalMO}, and the error of the Bayes-optimal learner \cite{opper1991}. In these models each output is a scalar, binary or real, so neither an embedding dimension nor a vocabulary appears.

Several recent works have built solvable models of transformer-like learning, each in a specific scaling regime. Attention-indexed models take the embedding dimension large at finite sequence length~\cite{Fab_attention_indexed}; theories of in-context regression take sequence length and dimension jointly large~\cite{bordelon_ict, Lu_ICT}; models of factual recall store associations between single tokens in a $d\times d$ matrix and send $d\to\infty$~\cite{nichaniICLR2025, giorlandino2026factual, barnfield2026factual}. In these models either the embedding dimension is sent to infinity, or the input is a single token, or there is no vocabulary; in none of them can the embedding dimension and the vocabulary size be varied at finite values. Other works study the attention structure of trained transformers in the Bayesian setting~\cite{sompo_mignacco}, regression from long sequences of high-dimensional tokens~\cite{maillard}, how networks acquire hierarchical grammars over a vocabulary~\cite{cagnetta2024structure}, or the dynamics of tokens under self-attention~\cite{gao2026cluster, gao2026repulsive}. 

In trained LLMs, among the sequence length $N$, the embedding dimension $d$ and the vocabulary size $D$, the sequence length is the largest, $N\sim 10^5$--$10^6$ tokens, while $d\sim 10^3$--$10^4$ and $D\sim 10^4$--$10^5$.
Motivated by this regime, we study a minimal model in which only the sequence length is taken large, $N\to\infty$ at fixed $\alpha=P/N$, since it is the largest of the three scales, while $d$ and $D$ are kept finite and arbitrary. The model is the perceptron extended to sequences of embedded tokens: a linear predictor that assigns a $d\times d$ matrix $\mathbf W_i$ to each position $i$ of the sequence, trained in the teacher--student setting~\cite{sompo_perc, watkin1993statistical, Engel2001StatisticalMO}. 
We call this the \textit{token perceptron}.
Taking $N\to\infty$ exploits the extensive number $Nd^2$ of parameters and allows a direct generalization of the replica theory of the perceptron~\cite{E_Gardner_1988, sompo_perc, Engel2001StatisticalMO, damico_ICLR}; the large-$d$ behavior is then derived from the finite-$d$ equations. The position-dependent weights are a simplification of attention, where positional information enters through encodings mixed with the token representations~\cite{NIPS2017_3f5ee243}.

To separate the role of the embedding dimension from that of the vocabulary, we consider two cases: continuous embeddings on the $d$-dimensional sphere and embeddings drawn from a vocabulary.
\begin{itemize}
    \item Continuous embeddings isolate the role of the embedding dimension. For these, we compute the generalization error for any rotationally invariant loss and find that the natural load is $\widetilde\alpha=P/(Nd)$. For $d>1$, a normalized label fixes the direction of the teacher output and therefore gives $d-1$ linear equations on the weights per example: classification becomes noiseless regression up to an overall scale. The teacher is recovered exactly above $\alpha_{\rm rec}=d^2/(d-1)$, the analogue of the interpolation threshold of the linear perceptron~\cite{krogh1992generalization}, and spherical cross-entropy reaches it in the limit of large inverse temperature. We also compute the storage capacity for random associations under two definitions of alignment. 
    \item Embeddings drawn from a discrete vocabulary are closer to the task of a transformer, which outputs one of $D$ tokens rather than a continuous vector. They also present a harder learning problem that does not reduce to regression: the label identifies only the Voronoi cell containing the teacher output, so each example constrains the weights through inequalities rather than equations. This makes the vocabulary case, rather than the continuous one, the natural extension of the classical perceptron. For this task, we compute the Bayes-optimal error for a given embedding matrix through an approximate spectral ansatz, and show that learning becomes anisotropic: the optimal learner first resolves the directions in which the vocabulary has the largest variance.
\end{itemize}
 
\begin{figure}[]
    \centering
    \includegraphics[width=0.99\linewidth]{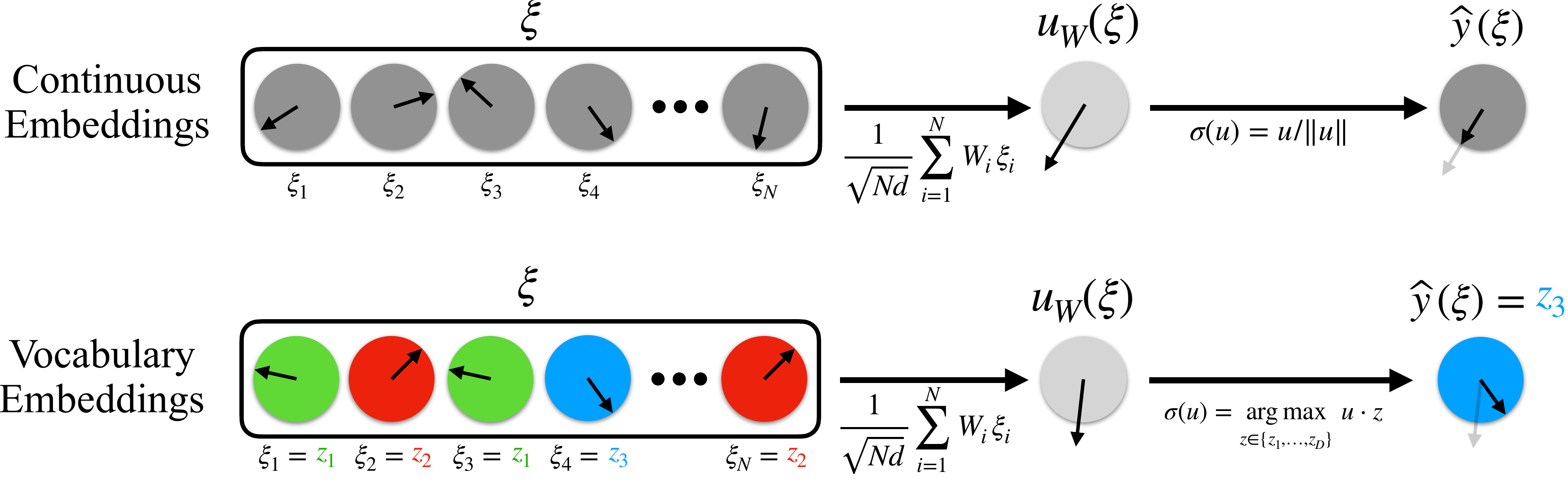}
    \caption{
    \textbf{Schematic representation of the learning problem with \emph{continuous} (top) and \emph{vocabulary} (bottom) token embeddings.}
    In both settings, the token perceptron maps the input sequence
    $\bm\xi=(\bm \xi_1,\ldots,\bm\xi_N)$ to the field
    $\bm u_W(\bm \xi)=(Nd)^{-1/2}\sum_{i=1}^{N}\mathbf W_i\bm \xi_i$.
    \textit{(Top)} continuous embeddings, with input vectors on the unit
    sphere $\mathbb S^{d-1}$ and prediction
    $\hat{\bm y}(\bm \xi)=\bm u_W(\bm \xi)/\|\bm u_W(\bm \xi)\|$.
    \textit{(Bottom)} vocabulary embeddings, with each input vector drawn
    from a fixed vocabulary $\{\bm z_1,\ldots,\bm z_D\}\subset\mathbb S^{d-1}$.
    The prediction is the vocabulary vector with maximal overlap
    with $\bm u_W(\bm \xi)$.
    }
    \label{fig:vocabulary_scheme}
\end{figure}

The paper is organized as follows. In Sec.~\ref{sec:quantities_general} we define the task and architecture that generalize the perceptron to multidimensional variables, together with the corresponding generalizations of the cross-entropy loss to continuous and vocabulary embeddings, that are two natural and valid possible extension of classical results. In Sec.~\ref{sec:continuous_embeddings} we solve the teacher-student problem for embeddings with continuous support (Fig.~\ref{fig:vocabulary_scheme} (top)), where each token embedding is constrained to the hypersphere $\mathbb{S}^{d-1}$ embedded in $d$ dimensions. This setting is motivated by normalization schemes widely used in modern language models, such as RMS normalization \cite{RMS}. In addition in Sec.~\ref{sec:mem_rand_input} we study memorization of random inputs, a task analogous to recent results on \textit{factual recall} \cite{nichaniICLR2025,giorlandino2026factual}. In Sec.~\ref{sec:vocabulary_learning} we study vocabulary classification (Fig.~\ref{fig:vocabulary_scheme} (bottom)), a multiclass supervised setting in which the geometry of a discrete set of vocabulary embeddings explicitly affects learning and generalization. This is a genuinely multidimensional task absent from the scalar perceptron setting and provides a simplified model of next-token prediction, the core objective of autoregressive LLMs \cite{radford2018improving}.

\section{Extension of classical quantities to the token perceptron}
\label{sec:quantities_general}

\subsection{Data: sequences of embedded tokens}
\label{sec:variables}

We consider a training dataset of $P$ input couples 
\begin{equation}
    \mathcal D=\{\boldsymbol{(\xi}^\mu, \bm y^\mu)\}_{\mu=1}^{P}, \quad \boldsymbol \xi^\mu \in \mathbb R^{N\times d}\,,\quad \bm y^\mu \in \mathbb R^d,
\end{equation}
where each $\boldsymbol \xi^\mu$ is a sequence of $N$ vectors embedded in a $d$-dimensional space $\bm \xi^\mu_i \in \mathbb R^d$ with unit norm, $\|\bm \xi^\mu_i\|^2=\sum_{\gamma=1}^d (\xi^\mu_{i, \gamma})^2=1$, and $\bm y^\mu$ is the corresponding label, with the same unit norm $\|\bm y^\mu\|^2=\sum_{\gamma=1}^d (y^\mu_{\gamma})^2=1$. 
The rule for assigning labels is specified in Sec.~\ref{sec:training_settings}. 
Each token $\bm \xi^\mu_i$ plays the role of an input variable of the classical perceptron, promoted from a scalar to a $d$-dimensional vector.
Token embeddings are drawn independently across positions and examples, uniformly from one of two supports.\footnote{Sampling tokens independently across positions is a standard assumption in solvable models of sequence and in-context learning \cite{Lu_ICT,bordelon_ict}, at variance with other results that aim to model sequential correlations \cite{letey2026sequential}. The model is consequently supervised rather than self-supervised: our aim is not to model the statistics of natural sequences, but to isolate how the geometry of their embedded tokens controls learning and generalization.}
We name the two cases after the support of the embedding, continuous or vocabulary embeddings. Throughout, “token” denotes an element of the input sequence, as in transformer architectures.

\paragraph{Continuous embeddings.} In this case, tokens and labels can be any vector on the unit hypersphere $\mathbb{S}^{d-1}$. This setting is related to inputs made of continuous-valued elements, as in Vision Transformers~\cite{dosovitskiy2021an}, in which each element is a patch of many pixels with approximately continuous RGB values.

\paragraph{Vocabulary embeddings.} This case is closer to data processed by Large Language Models, where tokens correspond to unique words or part of them \cite{sennrich2016token,kudo2018token}. In this case we introduce a vocabulary of tokens as their $d$-dimensional embedding representations
\begin{equation}
    \mathcal{Z}
    =
    \{\bm z_1,\ldots,\bm z_D\}
    \subset \mathbb{R}^{d},
\end{equation}
containing $D$ possible token embeddings. Equivalently, the vocabulary can be represented by the embedding matrix
\begin{equation}
    \mathbf Z =
    \begin{pmatrix}
        \bm z_1^\top\\
        \vdots\\
        \bm z_D^\top
    \end{pmatrix}
    \in \mathbb{R}^{D\times d}.
\end{equation}
Each vocabulary token $k\in\{1,\ldots,D\}$ is therefore associated with a continuous representation $\bm{z}_k\in\mathbb{R}^{d}$. {As for continuous embeddings, every vocabulary vector is normalized individually: $\|\bm{z}_k\|=1$.} An input sequence is obtained by drawing a token index at every position $i=1,\dots,N$,
\begin{equation}
    s_i \sim \mathrm{Unif}\{1,\ldots,D\}
    \qquad \rightarrow \qquad
    \bm \xi_i = \bm z_{s_i}
\end{equation}

Labels are also vocabulary vectors, selected by the rule of Sec.~\ref{sec:training_settings}.

The novelty of this vocabulary setting is that the geometry of the embedding matrix $\mathbf Z$ affects the learning process, meaning that two different vocabularies that share the same $d,D,\alpha$ define different learning problems. In this setting the vocabulary is quenched, meaning that it does not change during training. This is a simplification compared to real networks that train the embedding representations together with the rest of the network, however recent results have shown that even on large scale LLMs embeddings of words are represented in a mostly predictable and explainable way, meaning mostly independent of the particular setting \cite{Daniel_1,Daniel_2}, and then our approximation can be imagined as a fast converge of embeddings towards these learned representations, followed by a slower network learning. In addition, in our case the interest is in general properties of vocabulary classification, and therefore for simplicity we study vocabularies whose elements are extracted uniformly over the $d$-dimensional sphere, analogously to the continuous case. However our analytical results depend \emph{explicitly} on the fixed embedding matrix $ \mathbf Z$, and therefore could be applied directly to real embeddings.

\subsection{Task and training procedure}

\label{sec:training_settings}

We study the teacher-student setting \cite{sompo_perc}, where the labels
$\bm y^\mu$ are assigned by a one-layer neural network with weights
$\mathbf W^\star\in\mathbb R^{N\times d\times d}$, called \emph{the teacher}, through the rule
\begin{equation}
    \bm y(\bm \xi)=\sigma\!\left(\bm v(\bm \xi)\right),
    \qquad
    \bm v(\bm \xi)=\frac{1}{\sqrt{Nd}}\sum_{i=1}^{N}\mathbf W^\star_i\,\bm \xi_i,
\end{equation}
where $\mathbf W^\star_i\in\mathbb R^{d\times d}$ is the $i$-th weight matrix of the teacher and $\bm v(\bm \xi)\in \mathbb R^d$ is the teacher pre-activation, or field. The activation function $\sigma(\cdot)$ projects the field onto the closest vector in the embedding support, i.e.\ the set of values that inputs and labels can take. 

In the first setting (continuous embeddings) $\sigma$ normalizes the field to unit norm, $\sigma(\bm v)=\bm v/\|\bm v\|$, reducing to the classical perceptron activation $\operatorname{sign}(v)$ for $d=1$. In the second (vocabulary embeddings), it selects the vocabulary vector with maximal overlap with the field.

The teacher weights are sampled uniformly on the sphere $\|\mathbf W^\star\|_F^2=\sum_{i=1}^N \sum_{\gamma, \delta =1}^d (W^\star_{i, \gamma \delta})^2= Nd^2$ and fixed for a given training task. 

Another one-layer neural network, called \emph{student}, has the same architecture of the teacher and learnable weights $\mathbf W\in\mathbb R^{N\times d\times d}$ constrained to the same sphere $\|\mathbf W\|_F^2=Nd^2$. For a given input sequence $\bm \xi$, the student predicts an output using the same rule as the teacher:
\begin{equation}
    \hat{\bm y}(\bm \xi)=\sigma\!\left(\bm u(\bm\xi)\right),
    \qquad 
    \bm u(\bm\xi)=\frac{1}{\sqrt{Nd}}\sum_{i=1}^{N}\mathbf W_i\bm \xi_i,
    \label{eq:student}
\end{equation}
where $\mathbf W_i\in\mathbb R^{d\times d}$ is the $i$-th weight matrix of the student and $\bm u(\bm \xi)\in \mathbb R^d$ is the student pre-activation.

For vocabulary embeddings, 
the correct label $k^\mu$ of a given input $\bm{\xi}^\mu$ is the token closer to the prediction of the teacher (analogously $\hat k^\mu$ for the student prediction)
\begin{equation}
    \begin{gathered}
    k^\mu = \operatorname*{arg\,max}_{k}\bm z_k\cdot {\bm v^\mu},\qquad \bm y^\mu=\bm z_{k^\mu},\\
    \hat k^\mu=\operatorname*{arg\,max}_{k}\bm z_k\cdot\bm u^\mu,\qquad \hat{\bm y}^\mu=\bm z_{\hat k^\mu}
    \end{gathered}
    \label{eq:main_dict_label}
\end{equation}

The student is trained on the dataset $\mathcal{D}$ of Eq.~(1), and its weights are chosen to minimize the training loss
\begin{equation}
    \mathcal L(\mathbf W;\mathcal D)=\sum_{\mu=1}^{P}\ell\!\left(\bm{u}(\bm{\xi}_\mu),\bm y^\mu\right),
    \qquad
    \overline{\mathbf W} = \operatorname*{arg\,min}_{\mathbf W:\,\|\mathbf W\|_F^2=Nd^2}\;\mathcal{L}(\mathbf W;\mathcal D),
    \label{eq:minimzers}
\end{equation}
where $\ell(\cdot,\cdot)$ is a per-sample loss measuring the discrepancy between the student prediction and the teacher label.


\subsection{Exponential losses and learning rules for embeddings}
\label{sec:losses}
In order to define a training loss, we consider conditional distributions belonging to the exponential family
\begin{equation}
    dp(\bm{y}|\bm{u})
    =
    \frac{\exp\!\left(\lambda\,\bm{y}\!\cdot\!\bm u\right)}
    {\widetilde{Z}_{\varrho}(\bm u)}d\varrho(\bm{y}),
    \qquad
    \widetilde{Z}_{\varrho}(\bm u)
    =
    \int
    \exp\!\left(\lambda\,\bm x\!\cdot\!\bm u\right)
    d\varrho(\bm x),
    \label{eq:general_conditional}
\end{equation}
where $\bm u=\frac{1}{\sqrt{Nd}}\sum_{i=1}^{N}\mathbf W_i\bm\xi_i$
is the pre-activation of the network, $\bm y=\sigma(\bm v)$is the label and $\varrho(\cdot)$ is the measure specifying the space and prior geometry of the admissible outputs. The negative conditional log-likelihood, rescaled by the inverse temperature $\lambda$, is
\begin{equation}
    \ell(\bm{u}, \bm{y})
    =
    -\bm{y}\!\cdot\!\bm{u}
    +
    \lambda^{-1}\log \widetilde{Z}_{\varrho}(\bm{u})
    \label{eq:general_CE}
\end{equation}
This form arises naturally from pseudo-likelihood \cite{besag1974spatial,PL_potts,DAMICO2026131497}: in a Gibbs model, conditioning one variable on all the others removes the global partition function and leaves a local Gibbs distribution in which the conditioned variable interacts with an effective local field. For pairwise bilinear interactions this conditional distribution has precisely the form of Eq.~\eqref{eq:general_conditional}, with a normalization performed only over the admissible states of the single output variable. Pseudo-likelihood then replaces the generally intractable joint likelihood by a product of such locally normalized conditional probabilities.\\
\\Here we adopt this local conditional construction directly as a supervised learning rule, in what could be seen as the generalization of cross-entropy to multidimensional labels. In particular, the interaction term $-\bm y\!\cdot\!\bm u$ remains unchanged, while the measure $\varrho$, present in the normalization term $\widetilde{Z}_\varrho$, determines how the learning of the task is affected by embedding geometry. The discrete and continuous losses considered below can consequently be obtained as different realizations of the same conditional model:
\begin{enumerate}
    \item For fixed-norm continuous vectors $d\varrho$ is the uniform measure on the sphere and the conditional distribution is von Mises-Fisher, giving 
    \begin{equation}
        \ell_{\rm sph}=-\bm y\!\cdot\!\bm u+\lambda^{-1}\log K_d(\lambda\|\bm u\|)
        \label{eq:sphericalPL}
    \end{equation}
    where  
    \begin{equation}
        K_d(x)\equiv\int_{\mathbb S^{d-1}} d\Omega(\hat{\bm x})\,e^{x\,\hat{\bm x}\cdot\hat{\bm u}}=(2\pi)^{d/2}\,x^{1-d/2}I_{d/2-1}(x)
    \end{equation}
    with {$x=\lambda\|\bm u\|$} and $I_\nu$ the modified Bessel function of the first kind, used already as a training loss for a realistic setting in \cite{scott2021vMF}. In the following, we refer to the loss in Eq.~\eqref{eq:sphericalPL} equally as 'spherical pseudolikelihood' and 'spherical cross-entropy'.
    \item Restricting $\varrho$ to the vocabulary $\mathbf Z$ gives
    \begin{equation}
        \ell_{\mathcal Z} = -\bm y\!\cdot\!\bm u+\frac{1}{\lambda}\log\sum_{k=1}^{D}e^{\lambda \bm z_k\cdot\bm u}
        \label{eq:loss_dict}
    \end{equation}
    namely a softmax cross-entropy. The difference with usual cross-entropy with one-hot classes is that here the logits are fixed by the embedding geometry itself, and then correlations and distances between vocabulary vectors determine which classes compete with one another.
\end{enumerate}
Both quantities coincides with common binary cross-entropy (or logistic loss) when $d=1$, because in this limit the continuous normalization $K_d(\lambda\|\bm u\|) \rightarrow2\text{cosh}(\lambda\|\bm u\|)$, and if $D=2$ and $z_1=+1,z_2=-1$, we have again $\sum_{k=1}^{D}e^{\lambda z_k\cdot\bm u} = 2\text{cosh}(\lambda\|\bm u\|)$.


\section{Continuous spherical embeddings}
\label{sec:continuous_embeddings}

We first consider input and output embeddings with a continuous support, where each input sequence is made of $N$ vectors drawn independently and uniformly on the unit sphere, and the teacher is sampled uniformly from the sphere $\|\mathbf  W^\star\|_F^2=Nd^2$ and the student is constrained to the same sphere $\|\mathbf  W\|_F^2=Nd^2$. 
First in Sec.~\ref{sec:replica_cont} we will give a quick overview of the analytical methods that allow to predict the properties of the training loss minimizers. In Sec.~\ref{sec:gen_err_cont} we will use those results to study the generalization error of various learning rules. In Sec.~\ref{sec:continuous_exact_recovery} we show that, given enough data, there exists a transition where exact teacher recovery is possible. Finally in Sec.~\ref{sec:mem_rand_input} we study memorization capacity of random input-output association, where we identify and solve two possible generalization of storage property of classical perceptrons. \\
\subsection{Analytical overview for the continuous case}
\label{sec:replica_cont}

We want to characterize the properties of the minimizers of the training loss as a function of the training dataset size $\alpha$ and embedding dimension $d$.
For the continuous case, we will focus on losses of the form 
\begin{equation}
    \mathcal{L}= \sum_{\mu=1}^{\alpha N}V(
    \bm u^\mu\cdot\bm y^\mu,
    \|\bm u^\mu\|^{2}) =\sum_{\mu=1}^{\alpha N}V(
     h^\mu,
    \rho^\mu) \;,
\end{equation}
where $h^\mu=\bm u^\mu\cdot\bm y^\mu$ is the scalar product between the
pre-activation $\bm u^\mu$ of the student network and the teacher output
$\bm y^\mu$, and $\rho^\mu=\|\bm u^\mu\|^{2}$ is the squared norm of the
student pre-activation. The loss in Eq.~\eqref{eq:sphericalPL} belongs to this class.
The class of losses we consider extends that of the scalar case $d=1$,
\begin{equation}
    \mathcal{L}= \sum_{\mu=1}^{\alpha N}V(u^\mu \cdot \operatorname{sign}(v))=\sum_{\mu=1}^{\alpha N}V(\Delta^\mu) \, ,
\end{equation}
where $\Delta^\mu$ is usually called the stability~\cite{Engel2001StatisticalMO}, by adding a dependence on the squared norm $\rho^\mu$. We do this in $d>1$ since, in general, the student pre-activation takes the form
\begin{equation}
    \bm u^\mu = h^\mu\,\bm y^\mu + \bm u_\perp^\mu \, ,
\end{equation}
where $\bm u_\perp^\mu$ is a component perpendicular to the teacher output which does not affect the value of $h^\mu$, but whose presence we want to penalize.

In order to analytically describe the minimizers of this type of losses, we will use the non-rigorous replica method from the statistical mechanics of disordered systems \cite{mezard1987spin}; we only sketch the derivation here and refer to Appendix \ref{app:continuous_replica} for the details.
The main object we study is the replicated partition function, defined as 
\begin{equation}
    \overline{Z_\beta^n}
    =
    \int
    \prod_{a=1}^{n}d\varrho(\mathbf  W^a)
    \,
    \overline{
    \exp\left[
        -\beta
        \sum_{\mu=1}^{P}
        \sum_{a=1}^{n}
        V(h^{a,\mu},\rho^{a,\mu})
    \right]
    },
    \label{eq:app_replicated_Z_main}
\end{equation}
where $\beta$ is a parameter with the role of an inverse temperature, $\mathbf W^a$, $a=1,\dots,n$, are $n$ replicas of the student weights, $d\varrho(\mathbf W)$ is the measure over the student weights enforcing the spherical constraint $\lVert \mathbf{W}\lVert^2 = N d^2$, $h^{a,\mu}$ and $\rho^{a,\mu}$ are the quantities defined above evaluated for replica $a$, and the overline denotes the average over the realisations of the inputs $\bm \xi^\mu$ and the teacher.

In the replica computation, the averaged free energy at a given temperature is then defined as 
\begin{equation}
    f_\beta =-\frac{1}{\beta N} \,\lim_{n\to0}\partial_n\log\overline{Z_\beta^n}\, ,
\end{equation}
and in the thermodynamic limit $N \to \infty$, only a subset of student configurations contributes to $f_\beta$. This subset is characterized by specific values of the student-teacher overlap and the overlap between two student replicas, both of which are matrices in the $d>1$ case: 
\begin{align}
    R_{\alpha\beta}^{a}
    &=
    \frac{1}{Nd}
    \sum_{i=1}^{N}
    \sum_{\gamma=1}^{d}
    W^{a}_{i,\alpha\gamma}
    W^{\star}_{i,\beta\gamma},\\
    Q_{\alpha\beta}^{ab}
    &=
    \frac{1}{Nd}
    \sum_{i=1}^{N}
    \sum_{\gamma=1}^{d}
    W^{a}_{i,\alpha\gamma}
    W^{b}_{i,\beta\gamma}.
    \label{eq:app_R_def_main}
\end{align}
To simplify the analytical analysis, we work in the isotropic replica symmetric (RS) ansatz, i.e.,  we assume that the overlap matrices are of the form
\begin{equation}
    Q_{\alpha\beta}^{ab}
    =
    \begin{cases}
        \delta_{\alpha\beta},
        & a=b,
        \\
        q\,\delta_{\alpha\beta},
        & a\neq b,
    \end{cases}
    \qquad
    R_{\alpha\beta}^{a}
    =
    r\,\delta_{\alpha\beta}.
    \label{eq:app_RS_ansatz_main}
\end{equation}
In this case, each replica of the student is characterized by the same two scalars $(r,q)$, and the replicated partition function reduces to an integral over them (App.~\ref{app:continuous_replica}),
\begin{equation}
    \overline{Z_\beta^n}
    \asymp
    \int dq\,dr\,
    \exp\left\{
        Nn\left[
            G_S(q,r)+\alpha\, G_E(q,r)
        \right]
    \right\},
    \label{eq:GS_GE_main}
\end{equation}
where the \emph{entropic} term $G_S$ measures the log-volume of student weight configurations with overlaps $(q,r)$, and the \emph{energetic} term $G_E$ accounts for the contribution of the training patterns through the potential $V$. The free energy is then $f_\beta(r,q)=-\beta^{-1}\left[G_S(q,r)+\alpha\, G_E(q,r)\right]$. In order to find the values $(r_*,q_*)$ corresponding to the equilibrium teacher-student and replica overlaps in the thermodynamic limit, we extremise $f_\beta(r,q)$ respect to the two parameters, obtaining the saddle point equations
\begin{equation}
    \left. \frac{\partial f_\beta}{\partial q} \right|_{r_*,q_*}=0\,, \quad \left. \frac{\partial f_\beta}{\partial r} \right|_{r_*,q_*}=0\,.
\end{equation}
In the zero-temperature limit $\beta\to\infty$, the extremizers of the free energy correspond to the minimizers of the training loss, so the saddle-point equations in this limit characterize their overlap with the teacher. Solving them numerically yields the theoretical prediction $r(\alpha,d)$ for the student-teacher overlap of the training loss minimizers, as a function of the dataset size and the embedding dimension.

\subsection{Embedding regression generalization error and comparison of losses}
\label{sec:gen_err_cont}
The saddle-point equations determine the relation $r(\alpha,d)$ between the control parameters $\alpha,d$ and the student-teacher overlap value $r$. We now want to connect this quantity with the generalization performances of the student networks. For an unseen input, the cosine similarity between student and teacher outputs is defined as
\begin{equation}
    \widehat{\Delta}
    =
    \frac{\bm u\cdot\bm v}
    {\|\bm u\|\,\|\bm v\|}
    =
    \bm y\cdot\hat{\bm y}.
\end{equation}
We define the generalization error for continuous embeddings as
\begin{equation}
    \epsilon_g(r)
    =
    \frac{1}{2}
    \left(
        1-\left\langle\widehat{\Delta}\right\rangle
    \right)
    =
    \frac{1}{4}
    \left\langle
        \|\bm y-\hat{\bm y}\|^2
    \right\rangle,
    \label{eq:generror}
\end{equation}
where the brackets $\langle \cdot \rangle$ indicate the average on the population of unseen inputs. For a fixed value of the embedding dimension $d$, the average is only a function of the student-teacher overlap $r$, and its expression can be written explicitly (App.~\ref{app:continuous_replica}) as:
\begin{equation}
    \epsilon_g(r)
    =
    \frac{1}{2}
    \left[
        1
        -
        r\,\frac{A_d^2}{d}\,
        {}_2F_1
        \left(
            \frac12,\frac12;
            \frac d2+1;
            r^2
        \right)
    \right],
    \label{eq:continuous_generalization_error}
\end{equation}
where
$A_d=\sqrt{2}\, \frac{\Gamma\,\left(\frac{d+1}{2}\right)}{\Gamma\,\left(\frac d2\right)}
$
and ${}_2F_1(a, b; c; x)$ is the Gauss hypergeometric function\footnote{These are functions defined by the hypergeometric series ${}_2F_1(a, b; c; x)=\sum_{n=0}^{\infty} \frac{(a)_n (b)_n}{(c)_n}\frac{x^n}{n!}$, where $(a)_n=\prod_{k=0}^{n-1} (a+k)$.} of parameters $a=b=1/2$, $c=d/2+1$ and variable $x=r^2$. \\
This definition of the generalization error reduces for $d=1$ to the standard perceptron error 
\begin{equation}
    \epsilon_g(r)
    =
    \frac{1}{\pi}\arccos r.
\end{equation}
Combining the definition of the generalization error in Eq.~\eqref{eq:generror} and the student-teacher overlap obtained by solving the saddle point equations of Section~\ref{sec:replica_cont}, we can obtain the generalization error as a function of the load $\alpha$ and the embedding dimension $d$:  $\epsilon_g(r(\alpha, d))= \epsilon_g(\alpha, d)$.

\begin{figure}[ht!]
    \centering
    
    \includegraphics[width=0.65\linewidth]
    {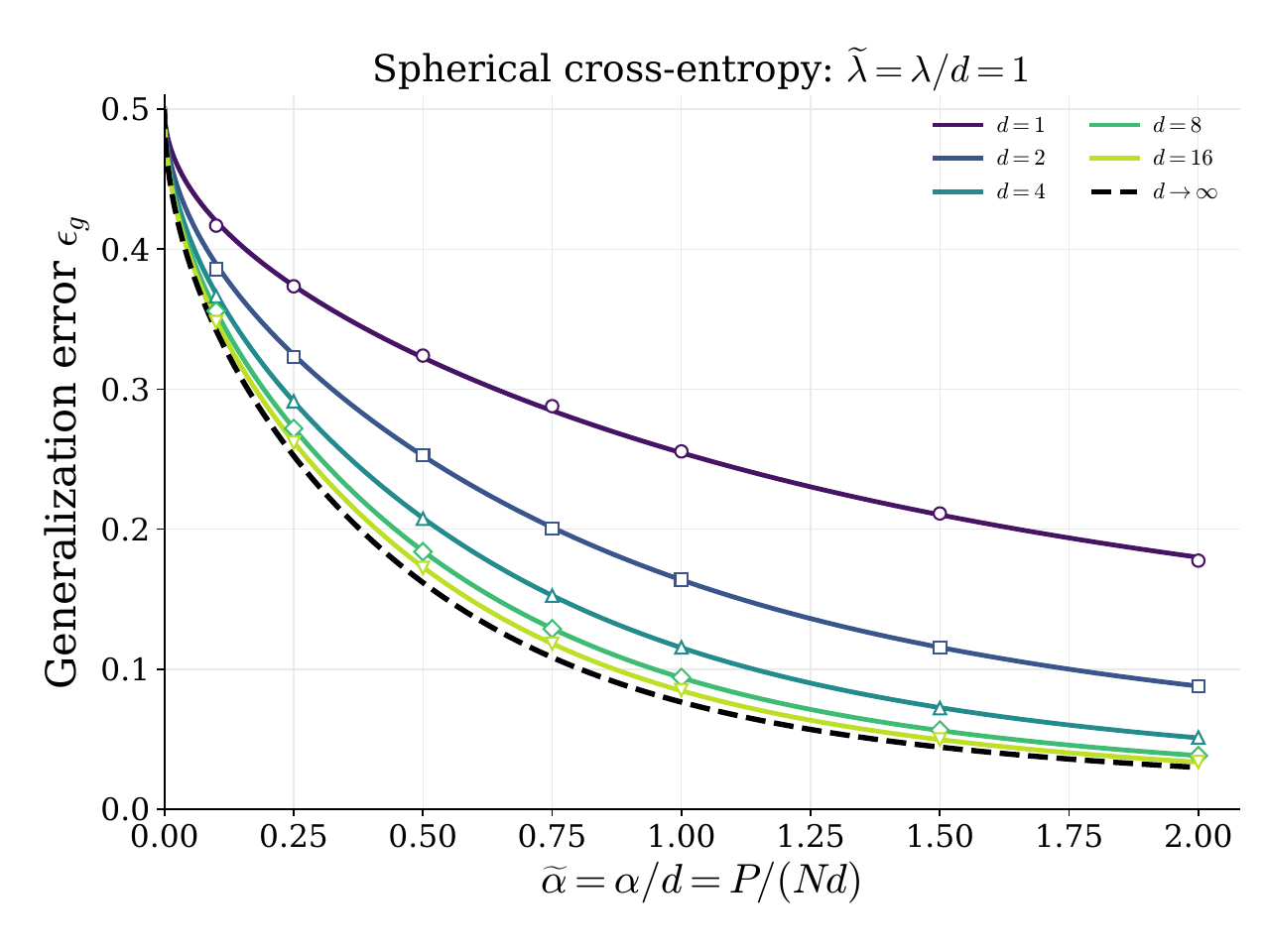}

    \caption{\textbf{Using the rescaled load, the generalization error $\epsilon_g$ curves converges for large embeding dimension.} We compare analytical predictions (solid lines) to numerical results (dots) for the generalization error $\epsilon_g$ of the spherical cross-entropy loss with $\tilde{\lambda}=\lambda/d = 1$ versus the rescaled load $\tilde{\alpha}$ for different values of the embedding dimensions $d$.}
    \label{fig:continuous_rs}
\end{figure}

Figure~\ref{fig:continuous_rs} compares the analytical predictions of the generalization error for the spherical pseudolikelihood loss in Eq.~\eqref{eq:sphericalPL}
with numerical simulations, where the student is obtained by optimizing the loss for a given realization of the problem using the L-BFGS-B optimization algorithm. In App.~\ref{app:continuous_replica}, we show that the analytical computation can be evaluated in the $N\gg d\gg 1$ regime; in this case, the load and the parameter $\lambda$ have to be rescaled accordingly
\begin{equation}
    \tilde{\lambda} = \lambda/d = O(1); \quad \tilde{\alpha}=\alpha/d =O(1),
\end{equation}
and to compare this with the finite-$d$ curves, we use these rescaled variables in the plot.

\subsection{Zero generalization error at finite load}
\label{sec:continuous_exact_recovery}

One interesting property of the token perceptron  is that for $d>1$ it becomes possible to achieve perfect recovery, i.e. zero generalization error, at a finite value of the load $\alpha$, even when a non-linearity $\sigma(\bm v)=\bm v/\|\bm v\|$ is applied to the teacher output to generate the examples labels. While this is known to happen in the linear perceptron case for $d=1$ \cite{hertz1989phase}, for the classical perceptron the Bayes-optimal error stays above zero for any finite value of $\alpha$ when $\sigma(v)= v/|v|=\operatorname{sign}(v)$ is used to construct the labels.
For $d>1$, requiring that the generalization error is zero corresponds to looking for a student weight configuration whose outputs have the same direction as the normalized teacher one, i.e. that the $d-1$ perpendicular components are zero, and that they lie on the correct half of the hyperplane: 
\begin{equation}
 (\mathbf I_d-\bm y^\mu\bm y^{\mu\top})\bm u^\mu=0,
 \qquad \bm y^\mu\cdot\bm u^\mu>0.
 \label{eq:continuous_recovery_constraints}
\end{equation}
Since $\bm u^\mu=\frac{1}{\sqrt{Nd}}\sum_{i=1}^{N}W_i\xi_i^\mu$ is linear in $\mathbf W$, each pattern correspond to $d-1$ linear
equations for the weights (corresponding to putting the $d-1$ components perpendicular to the teacher output $\bm v^\mu$ equal to zero). In the high-dimensional limit $N\to \infty$, for randomly extracted patterns, these conditions are independent. Since the number of student degrees of freedom is $Nd^2 -1$ (because of the spherical constraint), a simple counting argument gives:
\begin{equation}
 P(d-1)\geq Nd^2-1
 \quad\Longrightarrow\quad
 \alpha_{\rm rec}(d)=\frac{d^2}{d-1}.
 \label{eq:continuous_recovery_threshold}
\end{equation}
Above this load, the teacher is the only solution and the generalization
error is zero. The rescaled threshold is $\widetilde\alpha_{\rm rec}=\alpha_{rec}/d=2$ for $d=2$
and tends to $1$ as $d\to\infty$.

When $\tilde{\alpha}>\tilde{\alpha}_{rec}$ the spherical cross-entropy, Eq.~\eqref{eq:sphericalPL}, reaches this solution
in the limit $\lambda\to\infty$. Using $I_\nu(x)\simeq e^{x}/\sqrt{2\pi x}$ for
large $x$, one has $\lambda^{-1}\log K_d(\lambda\|\bm u\|)\simeq\|\bm u\|$, so that
\begin{equation}
    \ell_{\rm sph}\;\xrightarrow[\lambda\to\infty]{}\;
    \ell_\infty(\bm y,\bm u)=\|\bm u\|-\bm y\cdot\bm u
    =\|\bm u\|\left(1-\cos\theta\right)\geq 0 ,
    \label{eq:pl_infinite_lambda}
\end{equation}
where $\theta$ is the angle between $\bm u$ and $\bm y$. The leading term of the loss vanishes
only when $\bm u^\mu$ is parallel to $\bm y^\mu$, i.e.\ when the same constraints for zero generalization error
\eqref{eq:continuous_recovery_constraints} are satisfied. At large $d$ the
first correction to Eq.~\eqref{eq:pl_infinite_lambda},
$-\frac{d-1}{2\lambda}\log(\lambda\|\bm u\|)$, is of order $1/\widetilde\lambda$:
recovery then requires $\widetilde\lambda\to\infty$, i.e.\ $\lambda\gg d$.

\begin{figure}[ht!]

 \centering
 \includegraphics[width=0.65\linewidth]{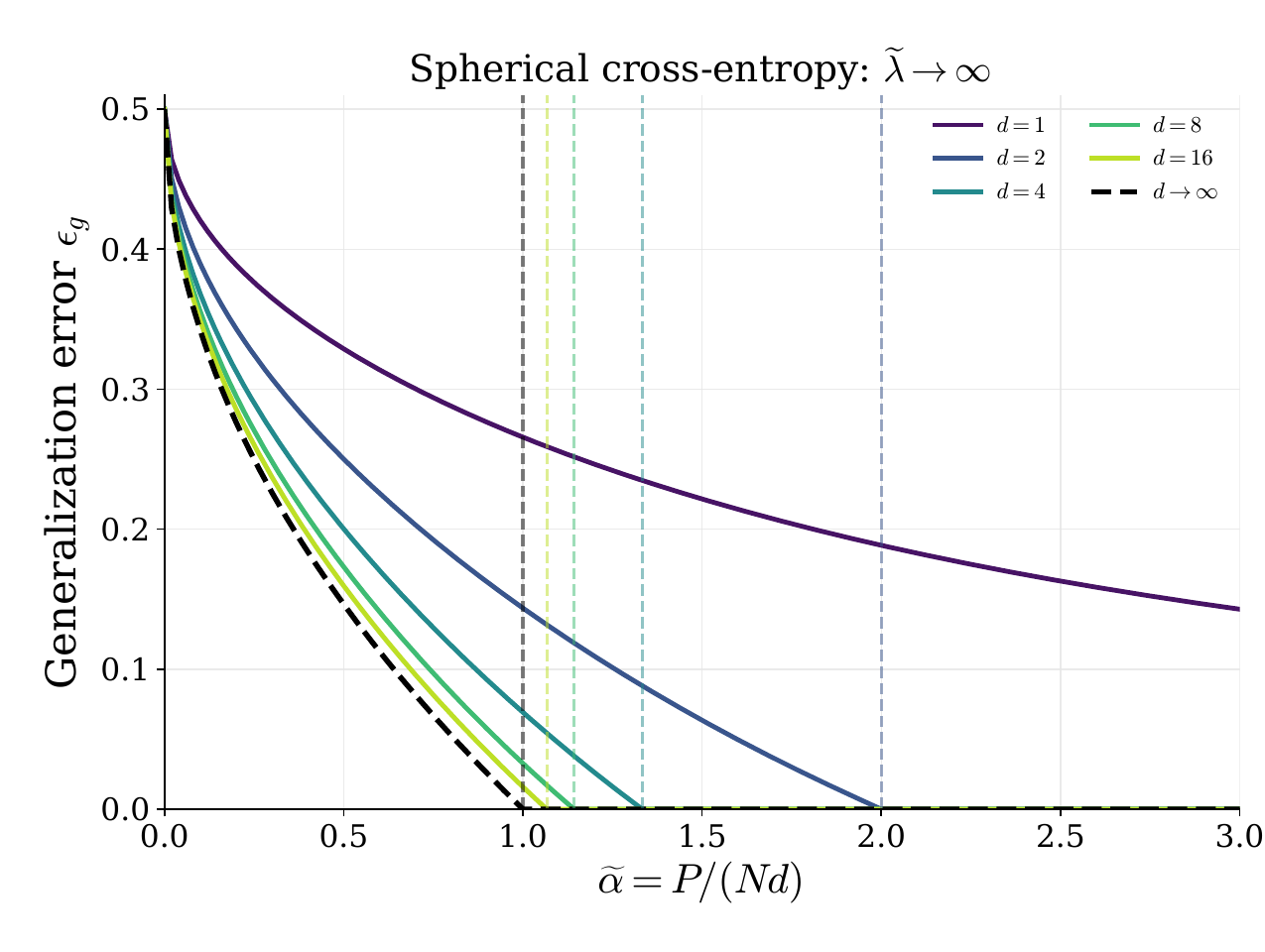}
 \caption{\textbf{Spherical cross-entropy learning curves at infinite $\widetilde\lambda$.} Generalization error $\epsilon_g$ versus $\widetilde\alpha=P/(Nd)$ for different embedding dimensions, obtained by minimizing cross-entropy at finite $\lambda$ and then taking $\widetilde\lambda\to\infty$. Vertical dashed lines mark the exact-recovery thresholds $\widetilde\alpha_{\rm rec}=d/(d-1)$ for $d>1$, Eq.~\eqref{eq:continuous_recovery_threshold}. The black dashed curve shows the $d\to\infty$ limit, with threshold $\widetilde\alpha_{\rm rec}=1$.}
 \label{fig:continuous_pl_infinite_lambda}
\end{figure}

Figure~{\ref{fig:continuous_pl_infinite_lambda}}
shows the generalization error curves for the cross-entropy loss as a function of
$\widetilde\alpha$ in the $\widetilde\lambda\to\infty$ limit. For $d>1$, the error reaches zero at $\widetilde\alpha_{\rm rec}=d/(d-1)$, marked by vertical dashed lines;
the $d=1$ maximum-margin curve remains positive at any finite load.

\begin{figure}[ht!]
 \centering
 \includegraphics[width=0.49\linewidth]{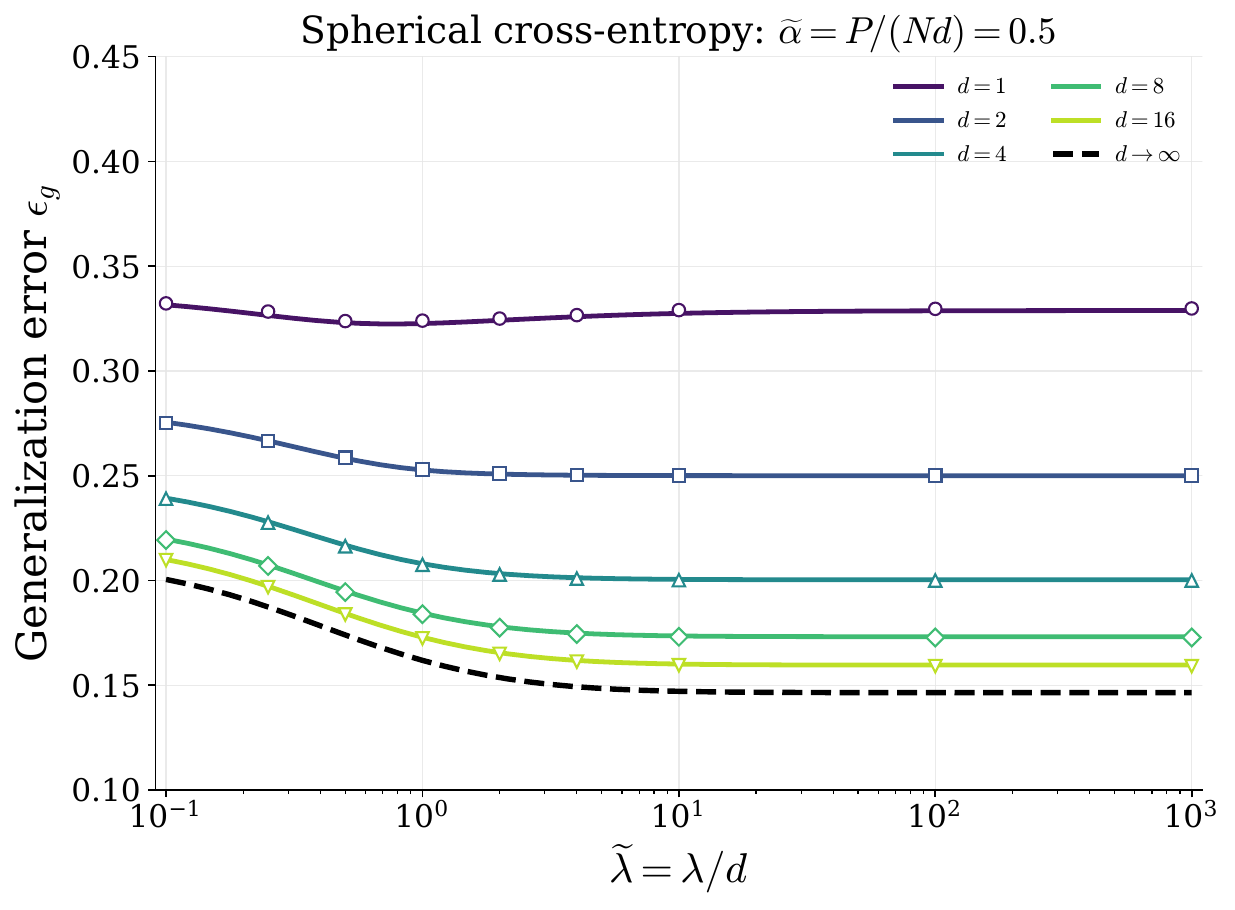}
 \hfill
 \includegraphics[width=0.49\linewidth]{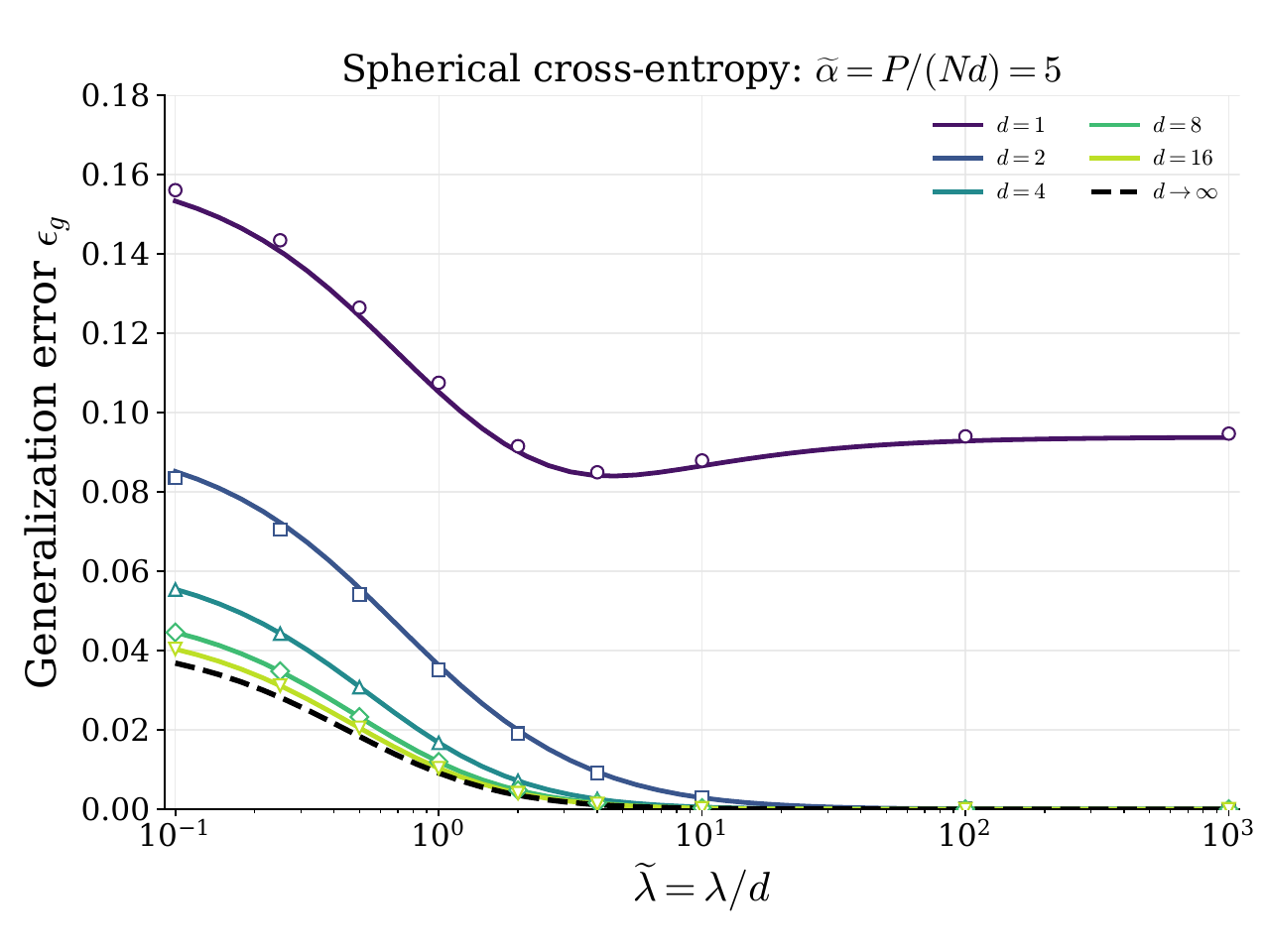}
 \caption{\textbf{For $d>1$, generalization error decreases as $\tilde{\lambda}$ increases and eventually reaches zero for $\tilde{\lambda}\to\infty$ for $\tilde{\alpha}>\frac{d}{d-1}=\tilde{\alpha}_{rec}(d)$.}
 Generalization error $\epsilon_g$ versus $\widetilde\lambda=\lambda/d$ at $\widetilde\alpha=0.5$ (left) and $\widetilde\alpha=5$ (right). At low load, all curves retain a positive error. At high load, cross-entropy recovers the teacher for every $d>1$, differently from the $d=1$ case where error stays finite.}
 \label{fig:continuous_lambda_recovery}
\end{figure}

Figure~\ref{fig:continuous_lambda_recovery} shows the generalization error as a function of the parameter $\tilde{\lambda}$ in two different regimes for different values of $d$.
At $\widetilde\alpha=0.5$ the load is smaller than the recovery threshold for all the dimensions, and all the curves retain a positive error at large
$\widetilde\lambda$. At $\widetilde\alpha=5$ the load is larger than the recovery threshold for all $d>1$ and the corresponding error tends to zero for $\tilde{\lambda}\to \infty$, while the $d=1$ curve converges to a positive value \cite{damico_ICLR}.

\subsection{Memorization of random inputs.}
\label{sec:mem_rand_input}
The same replica formalism can also be used to study a different question: how many \emph{random} input-output associations can be stored. This question is analogous to factual recall \cite{nichaniICLR2025, giorlandino2026factual, barnfield2026factual}; the difference here is the structure of input (a sequence of continuously embedded tokens instead of a single token extracted from a vocabulary) and the token perceptron (a tensor in $\mathbb{R}^{N\times d \times d}$ instead of a single matrix in $\mathbb{R}^{d\times d}$). By solving the saddle equations of Sec.~\ref{sec:replica_cont} imposing $r=0$ and by choosing a hard-constraint potential $V$, one can measure the Gardner volume of solutions satisfying random labels with a prescribed margin. The volume in the large $N$ limit corresponds to $e^{N \phi_*}$, where $\phi_*$ is the Gardner free entropy
\begin{equation}
\label{eq:free_entropy}
    \phi_* = \lim_{\beta\rightarrow\infty} \rm{extr}\{G_S(q_*, 0)+\alpha G_E(q_*, 0)\},\qquad \frac{\partial}{\partial q}G_S(q, 0)\Bigl |_{q=q_*} = -\alpha \frac{\partial}{\partial q}G_E(q, 0)\Bigl |_{q=q_*}
\end{equation}
where $G_S$ and $G_E$ are the entropic and energetic terms of Eq.~\eqref{eq:GS_GE_main} evaluated at $r=0$.

The volume of solutions vanishes when $q_*=1$, corresponding to the critical capacity
\begin{equation}
    \label{eq:critical_storage_capacity}
    \alpha_c = -\lim_{q_*\rightarrow 1_-} \frac{\frac{\partial}{\partial q}G_S(q, 0)\Bigl |_{q=q_*}}{\frac{\partial}{\partial q}G_E(q, 0)\Bigl |_{q=q_*}}.
\end{equation}
In this paragraph, we specialize in non-negative margins and convex sets of constraints, whose intersection gives rise to connected version spaces: we expect that in this setting, the system is RS, in analogy with the $d=1$ perceptron with spherical weights \cite{Engel2001StatisticalMO}. Given the potential $V(h, \rho)$, one can consider different capacity definitions that connect the storage problem to the geometry of network outputs (see Fig.~\ref{fig:scheme_minimization_V} in the Appendix for a scheme of the two different definitions).
\paragraph{Standard alignment (SA).}
The first case we present is the $d$-dimensional generalization of Gardner's original binary storage problem \cite{gardner1988optimalstorage}:
\begin{equation}
    \label{eq:potential_Gardner_standard}
    V_{\kappa}(h) = \Theta(\kappa-h)\Longrightarrow \lim_{\beta\rightarrow\infty} e^{-\beta V_{\kappa}} = \Theta(h-\kappa)
\end{equation}
The network output for a random input pattern satisfies the constraint if its coordinate $h$ in the random direction $\bm{y}$ exceeds a margin $\kappa$; for the remainder of this paragraph, we refer to this as \emph{standard alignment} (SA) problem. Therefore, each pattern imposes a single scalar constraint along the label direction, rather than the exact reproduction of the $d$-dimensional label. Since this choice of potential is independent of the norm $\rho$, the energetic part reduces to the scalar expression with standardized margin $\sqrt d\,\kappa$\;\footnote{In the storage setting $h=\bm{u}\cdot \bm{y}$, is $\mathcal{O}(1/\sqrt{d})$ by construction, while $t=\mathcal{O}(1)$ (see Eqs~\eqref{eq:app_h_pdf},\eqref{eq:app_chi_zeroT},\eqref{eq:app_relations_t_hbar_tbar} at $r=0$ in Appendix \ref{app:continuous_replica}). Thus, to isolate the action of $d$, to get eq. \eqref{eq:continuous_gardner_capacity} we wrote the constraints as $\tilde{h}>\sqrt{d} \kappa$, with $h=\tilde{h}/\sqrt{d}$. Note that in the VA case, instead, the inequality is invariant under a common rescaling of $h$ and $t$: $h>\kappa_{\rm eff} t \iff \widetilde{h}>\kappa_{\rm eff}\widetilde{t}$, with $\widetilde{t}=\sqrt{d}t$. 
}. By applying formula \eqref{eq:critical_storage_capacity} to SA \eqref{eq:potential_Gardner_standard} at $r=0$, we obtain
{\begin{equation}
    \alpha_c^{(\mathrm{SA})}(\kappa;d)
    =
    \frac{
        d^2
    }{
        \displaystyle
        \int_{-\infty}^{\sqrt d\,\kappa}
        Dt\,
        (t-\sqrt d\,\kappa)^2
    }
    =
    d^2\alpha_c^{\rm (G)}(\sqrt d\,\kappa),
    \label{eq:continuous_gardner_capacity}
\end{equation}}
where $\alpha_c^{\rm (G)}(\kappa)$ is the Gardner capacity of the spherical perceptron with binary patterns and
\begin{equation}
    Dt
    =
    \frac{dt}{\sqrt{2\pi}}e^{-t^2/2}.
\end{equation}
In particular $\alpha_c^{\rm (SA)}(0;d)=2d^2$, i.e.\ two scalar constraints per parameter, as for the perceptron, while $\alpha_c^{(SA)}\approx d/\kappa^2$ for $\kappa\gg 1/\sqrt{d}$. The increase in capacity simply reflects the augmentation of the weights space, with $Nd^2$ parameters.


\paragraph{Vector alignment (VA).} The second storage problem we present measures the volume of outputs aligned with the random labels within a cone of angle $\arccos{\kappa}$, where $0<\kappa<1$ has now the role of margin: recalling that $\frac{\bm{u}}{\lVert\bm{u}\rVert}\cdot \bm{y} \equiv \frac{h}{\sqrt{\rho}}$, the potential we use is
\begin{equation}
    \label{eq:potential_Gardner_angular}
    V_{\kappa}(h, t) = \Theta\left(\kappa-\frac{h}{\sqrt{\rho}}\right)\Longrightarrow \lim_{\beta\rightarrow\infty} e^{-\beta V_{\kappa}} = \Theta\left(\frac{h}{\sqrt{\rho}}-\kappa\right).
\end{equation}
With this definition, transverse directions in the student output enter the storage problem: we name this storage problem \emph{vector alignment} (VA) problem. Note that, at variance with SA, VA has no meaningful $d\rightarrow 1$ limit: it measures a genuine $d>1$ effect. We found the following critical load:
\begin{equation}
\begin{gathered}
    \label{eq:continous_alignment_capacity}
    \alpha_c^{\rm (VA)}(\kappa;d)
    =
    \frac{
        d^2
    }{
        \displaystyle
        \mathbb E_{
            z\sim\mathcal N(0,1),
            \,
            \bar t\sim\rm{CHI}_{d-1}
        }
        \left[
            D_\kappa^2(z,\bar t)
        \right]
    } \\ \\ \text{with} \\ \\
    D_\kappa^2(z,\bar t)
    =
    \frac{(z-\kappa_{\rm eff}\bar{t})^2}{1+\kappa_{\rm eff}^2}\Theta(\kappa_{\rm eff}\bar{t}-z)\Theta\left(z+\frac{\bar{t}}{\kappa_{\rm eff}}\right)+(z^2+\bar{t}^2)\Theta\left(-z-\frac{\bar{t}}{\kappa_{\rm eff}}\right)
\end{gathered}
\end{equation}

\begin{figure}[t]
    \centering
    \includegraphics[width=0.95\linewidth]{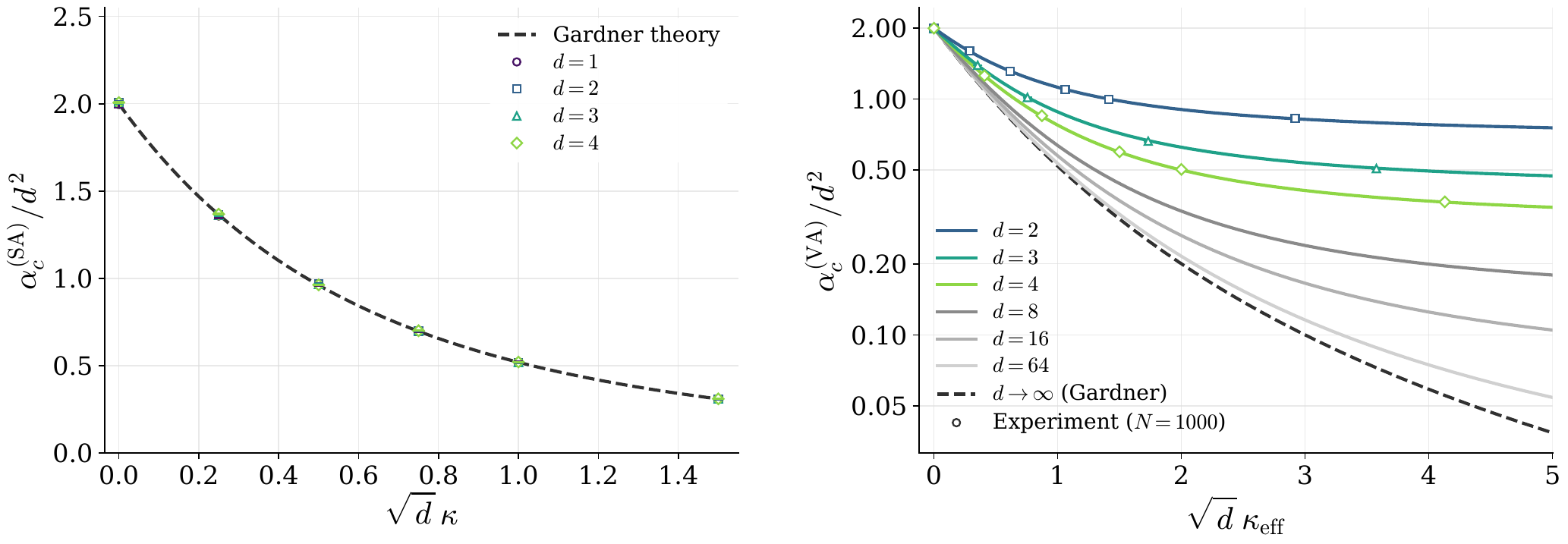}

    \caption{\textbf{Memorization capacity of random associations.}
    (Left) Critical load per degree of freedom $\alpha_c/d^2$ for standard alignment, Eq. \eqref{eq:continuous_gardner_capacity}. The curve shows the Gardner prediction in Eq.~\eqref{eq:continuous_gardner_capacity}. Markers show simulations at $d=1,2,3,4$ with $N=1000$.  
    (Right) Critical load per degree of freedom for vector alignment, Eq. \eqref{eq:continous_alignment_capacity}. Theoretical curves for $d=2, 3, 4, 8, 16, 64$ (continuous lines) and the $d\rightarrow\infty$ limit of eq. \eqref{eq:alpha_align_large_d} (dashed curve) are shown together with numerical simulations for $d=2, 3, 4$ of finite systems of size $N=1000$. 
    }
    \label{fig:continuous_memorization}
\end{figure}

where we introduced the effective margin $\kappa_{\rm eff} = \frac{\kappa}{\sqrt{1-\kappa^2}}$ and $\rm{CHI}_{d-1}(\cdot)$ is the chi-distribution with $d-1$ degrees of freedom.
As in Eq.~\eqref{eq:continuous_gardner_capacity}, we have $\alpha_c^{\rm (VA)}(0; d) = 2 d^2$, but for $\kappa_{\rm eff}>0$ the load per degree of freedom does not collapse on the $d=1$ Gardner's load as above: the misalignment of the network output affects the critical capacity in a non-trivial way. In particular, for $\kappa_{\rm eff}=\mathcal{O}(1/\sqrt{d})$ the capacity is $\mathcal{O}(d^2)$, but it becomes $\mathcal{O}(d)$ for $\kappa_{\rm eff}=\mathcal{O}(1)$: for instance, at $\kappa=1/\sqrt{2}$ ($\kappa_{\rm eff}=1$) its value is $\alpha_c^{\rm (VA)}(1/\sqrt{2}; d) = 2d$ and $\alpha_c^{\rm (VA)}(\kappa; d)\rightarrow \frac{d^2}{d-1/2}$ for $\kappa\rightarrow 1_-$ (corresponding to $\kappa_{\rm eff}\rightarrow \infty$).
In the large $d$ limit, the chi-distribution $\rm{CHI}_{d-1}$ in Eq.~\eqref{eq:continous_alignment_capacity} peaks at $\sqrt d$, and the student field has a typical cosine $\mathcal{O}(1/\sqrt{d})$ with any of the random teacher outputs: at leading order in large $d$, the effect of the transverse component is thus negligible, and by setting $\kappa_{\rm eff} = \frac{\widehat{\kappa}}{\sqrt{d}}$, we recover Gardner's formula as a scaling form:
\begin{equation}
\label{eq:alpha_align_large_d}
    \alpha_c^{\rm (VA)}(\hat{\kappa}; d) \approx d^2 \alpha_c^{\rm (G)}(\hat{\kappa}).
\end{equation}
The full derivation of the load formula of Eqs.~\eqref{eq:continuous_gardner_capacity} and \eqref{eq:continous_alignment_capacity} is shown in App.~\ref{app:continuous_classification_capacity}.

We tested the SA (Eq.~\eqref{eq:continuous_gardner_capacity}) and VA (Eq.~\eqref{eq:continous_alignment_capacity}) theoretical predictions with numerical simulations at $N=1000$.
In Fig.~\ref{fig:continuous_memorization}, we show the critical loads of eqs \eqref{eq:continuous_gardner_capacity}, \eqref{eq:continous_alignment_capacity} per degree of freedom $\alpha_c^{\rm (SA)}/d^2$ and $\alpha_c^{(\rm VA)}/d^2$, as functions of the rescaled projection margin $\sqrt d\,\kappa$ and the rescaled effective angular margin $\hat{\kappa} = \sqrt{d}\kappa_{\rm eff}$ respectively. Numerical simulations confirm that the rescaled SA capacity is independent of $d$ at fixed $\sqrt d\,\kappa$, reflecting the $Nd^2$ weight parameters and the single scalar constraint imposed by each pattern. On the other hand, simulations confirm that the VA capacity \eqref{eq:continous_alignment_capacity} has a residual dependence on $d$, quantifying the effect of orthogonal fluctuations.

The scaling of the storage of our model is enhanced by the larger number of degrees of freedom. At first sight, this result may seem trivial, however we stress that the information itself that has to be stored, corresponding to input sequence and output label, also increases with $d$, and in other similar memorization settings capacity scales slower with $d$ or even decreases. The Vector Hopfield network \cite{nicoletti2025statistical} has Hebbian $d\times d$ weights, thus still $Nd^2$ total number of weights per neuron, but its static\footnote{We define as static storage capacity the storage of patterns as local minima in an energy landscape, while dynamic storage as the load of patterns that can be retrieved by early stopping the dynamics \cite{nicoletti2025statistical, clark2026transient}.} storage capacity scales as $\alpha_c \propto 1/d $, while its dynamic storage capacity $\alpha_c \propto d$. Vector hopfield networks generalized to Riemann manifolds instead have a static capacity growing with $d$ \cite{galitski2026highcapacitygeneralizedhopfieldnetworks}.
Finally, factual recall associative memories \cite{giorlandino2026factual, barnfield2026factual} have a critical load scaling as $\alpha_c \propto \frac{d^2}{\log d}$.


\section{Vocabulary classification}
\label{sec:vocabulary_learning}

Section~\ref{sec:continuous_embeddings} showed that for $d>1$ continuous labels turn the problem into a noiseless regression: each example fixes the direction of the teacher output, and above the recovery threshold the teacher is the only solution. A vocabulary removes this mechanism. The label only states which Voronoi cell contains the teacher output, while the position of the output inside the cell is never observed, so each example constrains the weights through inequalities rather than equations. This is the multidimensional analogue of the sign nonlinearity of the classical perceptron, which is recovered for a vocabulary of two antipodal tokens. The relevant question is then how the geometry of the cells, fixed by the vocabulary, determines which information about the teacher each label reveals.

In this section we study the role of a \textit{fixed vocabulary} in the teacher-student setting. As explained in Sec.~\ref{sec:variables}, we have a vocabulary of $D$ tokens, embedded as $d$-dimensional vectors $\{\bm z_k\}_{k=1,\dots,D}\in \mathbb{R}^d$, and we define the embedding matrix as
\begin{equation*}
    \mathbf{Z} =  \begin{pmatrix}
        \bm z_1^\top\\  \vdots\\ \bm z_D^\top
    \end{pmatrix}
    \in \mathbb{R}^{D\times d}.
\end{equation*}
The vocabulary embedding vectors determine a set of Voronoi cells, and the geometry of such cells affects the learning problem, changing which directions of the teacher are easier or more difficult to learn.
We develop a theory \emph{explicitly} dependent on $\mathbf Z$ and therefore directly applicable to more complicated distributions or even real word embeddings \cite{Daniel_1,Daniel_2}. However we study learning curves and properties of the learned students for a vocabulary of unit vectors extracted from the uniform distribution over the $d$-dimensional sphere, in analogy with the previous case
\begin{equation}
    \bm{z}_k \sim \text{Uniform}(\mathbb{S}_{d-1}), \qquad ||\bm{z}_k||=1
\end{equation}
In order to isolate the effect of the fixed vocabulary we are interested in the properties of the statistically optimal predictor, at variance with the previous case in which we were interested in the role of the loss function, and therefore we set from the beginning the computation for the Bayes-Optimal case, following \cite{opper1991, Engel2001StatisticalMO}. We define the indicator function of the $k$-th decision cell as
\begin{equation}
    \Psi_k(\bm w;\mathbf Z) = \prod_{\ell\neq k} \Theta \left[(\bm z_k-\bm z_\ell)\cdot \bm w\right].
    \label{eq:main_dict_psi}
\end{equation}
that indicates whether $k$ is or is not the predicted token
\begin{equation}
    \Psi_k(\bm w;\mathbf Z)=\begin{cases}1,& \text{if }k=\displaystyle\operatorname*{arg\,max}_{j}
        \bm z_j\cdot \bm w, \\[2mm] 0,
        & \text{otherwise.}
    \end{cases}
\end{equation}
Using this we can write the Bayes posterior partition function as
\begin{equation}
    \mathcal Z(\mathcal D, \mathbf Z) = \int d\varrho(\mathbf W) \prod_{\mu=1}^{P}    \Psi_{k^\mu} \left( \bm u_W^\mu; \mathbf Z \right).
    \label{eq:main_dict_Z}
\end{equation}
that consists in the volume of all students $\mathbf W$ that classify correctly the training set, where $d\varrho(\mathbf W)$ is the uniform measure on $\|\mathbf W\|_F^2=Nd^2$ of the spherical teacher prior. In Appendix \ref{app:vocabulary_replica} we show the full derivation of the quenched average with replica method in the thermodynamic limit $N\rightarrow\infty$. The replica computation follows the same formal steps as in Sec.~\ref{sec:continuous_embeddings}, but the finite vocabulary changes the structure of the order parameters. In the continuous case the inputs have zero mean and covariance proportional to the identity, while here instead, given the realization of the vocabulary $\mathbf Z$, we have
\begin{equation}
    \mathbb E[\bm \xi_i] =\overline{\bm z}_Z = \frac{1}{D} \sum_{k=1}^{D}\bm z_k,
    \qquad  \operatorname{Cov}(\bm \xi_i) = \mathbf \Sigma_{\mathbf Z} = \frac{1}{D} \sum_{k=1}^{D} (\bm z_k-\overline{\bm z}_Z) (\bm z_k-\overline{\bm z}_Z)^\top .
    \label{eq:main_dict_mean_cov}
\end{equation}
Therefore, repeating the same large-$N$ central-limit argument used in the continuous case, the replicated pre-activations are Gaussian variables with mean
\begin{equation}
    \bm b^a
    =
    \frac{1}{\sqrt{Nd}}
    \sum_{i=1}^{N}
    \mathbf W_i^a\overline{\bm z}_Z
    \label{eq:main_dict_bias}
\end{equation}
and covariance
\begin{equation}
    \mathcal C^{ab}[\mathbf Z]
    =
    \frac{1}{Nd}
    \sum_{i=1}^{N}
    \mathbf W_i^a
    \mathbf \Sigma_Z
    \mathbf W_i^{b\top}.
    \label{eq:main_dict_exact_Q}
\end{equation}
and in general is not proportional to the identity given that the size of vocabulary $D$ is finite, even if the original distribution of tokens embeddings is isotropic. We assume that the bias in Eq. \eqref{eq:main_dict_bias} is learned exactly at every fixed $\alpha>0$, and therefore in what follows we do not treat $b^a$ anymore as an order parameter and all results will be derived for the centered quantities. This assumption is motivated by the fact that the bias shifts every output in the same direction and therefore can be inferred without reconstructing the sequence-dependent input-output mapping.\\
\\This full problem involves all embedding directions, meaning that in the end the solution would require a joint optimization of a $d\times d$ matrix saddle. Even though the full solution is in principle doable, we want a simpler description of the solution that may offer a more insightful explanation of the learning process. We find that this role could be played effectively by the \textit{spectral anisotropic Ansatz}, which consists in reducing the problem as diagonal in the basis of vocabulary covariance, resulting in $d$ order parameters that control learning mode-by-mode in this basis, that requires a joint optimization of a $d$-dimensional vector saddle. In Sec.~\ref{sec:anisotropic_main} we show a sketch of the replica derivation, in \ref{sec:scaling_curves} we compare learning curves and derive the natural scalings of parameters, and in \ref{sec:spec_results} we analyze how modes are learned differently depending on the regime of parameters.

\subsection{Analytical overview using a spectral anisotropic approximation}
\label{sec:anisotropic_main}
\begin{figure}[H]
    \centering        \includegraphics[width=0.48\linewidth]{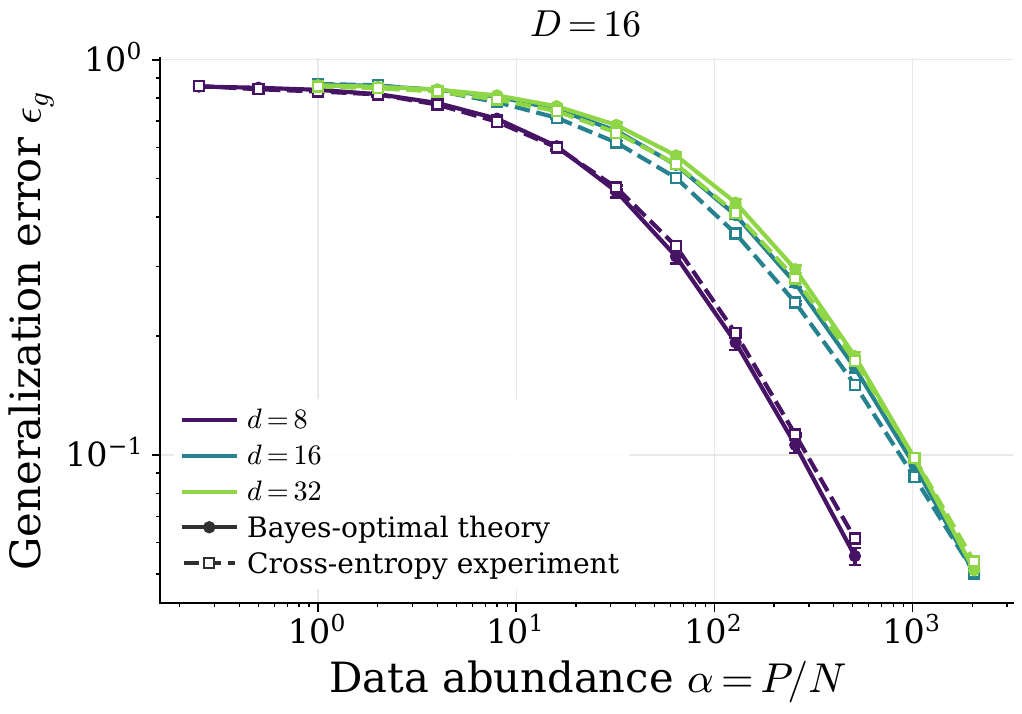}
    \includegraphics[width=0.48\linewidth]{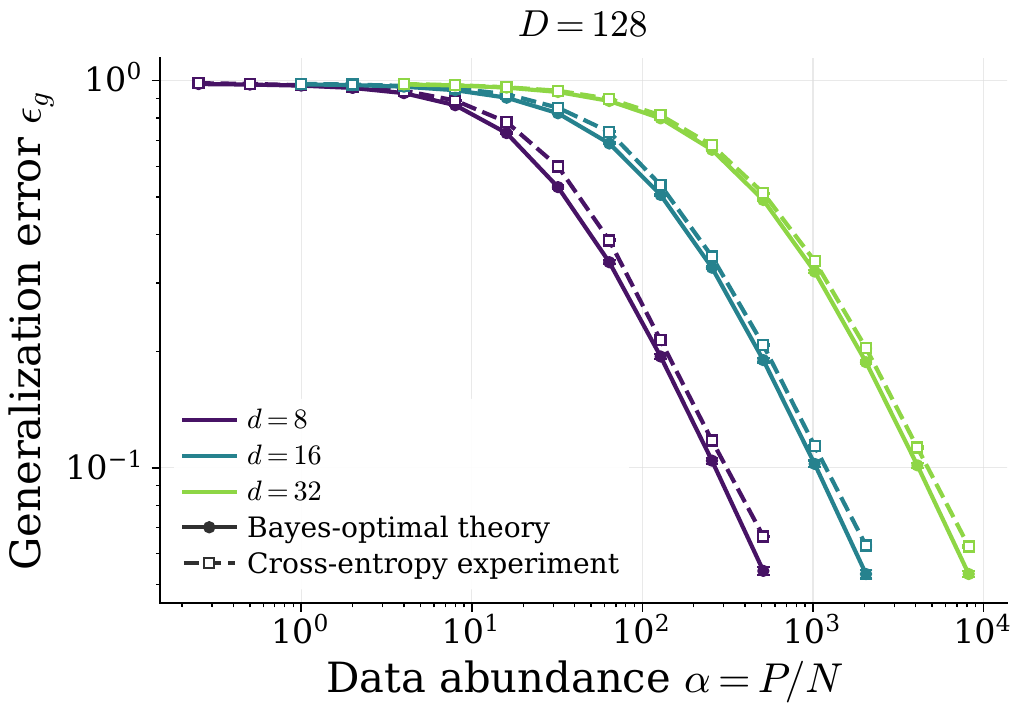}
    \caption{\textbf{Learning curves of generalization error for vocabulary classification task.} We compare the approximate Bayes-optimal analytical learning curves derived with the spectral anisotropic replica computation with real students trained minimizing cross-entropy loss without constraints on weights norm. The two minimization procedures are different in principle, but the fact that they are close imply that cross-entropy is able to find a solution with $\epsilon_g$ close to the optimal, similarly to \cite{aubin_convex_opt,damico_ICLR}. Both results are explicitly dependent on the extracted vocabulary $Z$, and therefore we generate 16 different random spherical vocabularies for each $d,D$ couple and we obtain prediction and numerical curves by averaging the error over the same 16 vocabularies. A numerical simulation with increasing number of variables $N$ would have the Bayes-optimal predictions increasingly becoming lower bounds for $\epsilon_g$. We note that when $d<D$ numerical results have higher $\epsilon_g$ then predictions as expected, while instead when $d\geq D$ some of the numerical points have lower $\epsilon_g$. This is compatible with the fact that our computation is an approximation of the full $d\times d$ problem, that we expect to work best in the regime $d<D$.}
    \label{fig:iso_vs_aniso}
\end{figure}

We retain the anisotropy of the vocabulary by working in the eigenbasis of its centered covariance,
\begin{equation}
    \boldsymbol{\Sigma}_Z=\mathbf{O}\boldsymbol{\Lambda} \mathbf{O}^\top,
    \qquad
    \mathbf\Lambda=\operatorname{diag}
    (\lambda_1,\ldots,\lambda_{\mathrm{r}_\Sigma}),
    \qquad
    \mathrm{r}_\Sigma=\operatorname{rank}(\mathbf \Sigma_Z)=\min(d,D-1)
    \label{eq:main_dict_spectrum}
\end{equation}
where only positive eigenvalues are retained. We assume symmetry between replicas $(a,b)$ and approximate the matrix of order parameters as diagonal in this basis, therefore we have a different replica overlap parameter $q_m$ for each spectral mode
\begin{equation}
    \left(\mathbf O^\top \mathbf Q^{ab}\mathbf O\right)_{m\ell}
    =
    \kappa_Z\,\delta_{m\ell}
    \left[
        \delta_{ab}
        +(\beta_Z+\gamma_Z q_m)(1-\delta_{ab})
    \right],
    \qquad m,\ell=1,\ldots,\mathrm{r}_\Sigma
    \label{eq:main_dict_spectral_ansatz}
\end{equation}
where $\kappa_Z$ is the total single-replica pre-activation field variance, $\beta_Z$ is the factor due to the already learned common bias, and $\gamma_Z=1-\beta_Z$ accounts for the fluctuating fraction. The value $q_m$ measures the overlap between the fluctuating fields of different replicas along vocabulary direction $m$: $q_m=0$ means that only their common bias is shared, whereas $q_m=1$ means that their fields coincide along that direction. The replicated partition function has the same structure as in the continuous case, Eq.~\eqref{eq:GS_GE_main},
\begin{equation}
    \overline{\mathcal Z^n}
    \asymp
    \int\prod_{m=1}^{\mathrm{r}_\Sigma}dq_m\,
    \exp\left\{
        Nn\left[
            G_S^{\mathrm{spec}}(\bm q)
            +\alpha G_E^{\mathrm{spec}}(\bm q; \mathbf Z)
        \right]
    \right\},
    \label{eq:main_dict_spectral_replicated_Z}
\end{equation}
but its saddle now determines the vector $\bm q=(q_1,\ldots,q_{\mathrm{r}_\Sigma})$. In particular, the entropic term $G_S^{\mathrm{spec}}$ measures the volume of students with the prescribed overlaps
\begin{equation}
    G_S^{\mathrm{spec}}(\bm q)
    =\frac{\mathrm{r}_Z}{2}\sum_{m=1}^{\mathrm{r}_\Sigma}
    \left[\log(1-q_m)+q_m\right],
    \qquad \mathrm{r}_Z=\operatorname{rank}(\mathbf Z)=\min(d, D).
    \label{eq:main_dict_spectral_GS}
\end{equation}
The energetic term $G_E^{\mathrm{spec}}$ instead accounts for student compatibility with the observed labels and therefore depends explicitly on the fixed vocabulary $\mathbf Z$, a feature specific to the presence of vocabulary embeddings. To understand its structure, we first rewrite the overlaps in Eq. \eqref{eq:main_dict_spectral_ansatz} in terms of centered Gaussian field
\begin{equation}
    \bm u \propto
    \sum_{m=1}^{\mathrm{r}_\Sigma}\bm O_m
    \left[
        \sqrt{q_m}\, h_m+\sqrt{1-q_m}\,  x_m
    \right],
    \qquad
    \bm h,\bm x\sim\mathcal N(0,I_{\mathrm{r}_\Sigma})
    \label{eq:w_spectral}
\end{equation}
where $\bm O_m$ is the $m$-th eigenvector of $\mathbf \Sigma_Z$, $h$ fields are shared by teacher and students, while $x$ are the individual posterior fields. For a fixed realization of $\bm h$, the posterior probability of token $k$ is
\begin{equation}
    p_k(\bm h, \mathbf  Z;\bm q)
    =
    \mathbb P_{\bm x}
    \left[
        \operatorname*{argmax}_{j=1,\ldots,D}
        \bigl(\bm z_j\cdot \bm u(\bm h, \bm x;\bm q)\bigr)
        =k
    \right].
    \label{eq:main_dict_spectral_pk}
\end{equation}
For a given realization of the fields $\bm h$, the probabilities $p_k$ describe how the residual uncertainty over the individual fields $\bm x$ is distributed among tokens. The energetic contribution corresponds to the negative conditional entropy of this token distribution with respect to $h$
\begin{equation}
    G_E^{\mathrm{spec}}(\bm q; \mathbf  Z)
    =
    \mathbb E_{\bm h}
    \left[
        \sum_{k=1}^{D}
        p_k(\bm h, \mathbf  Z;\bm q)
        \log p_k(\bm h, \mathbf  Z;\bm q)
    \right]
    =
    -H_Z(\bm q),
    \label{eq:main_dict_spectral_GE}
\end{equation}
Thus $H_Z$ measures the remaining uncertainty about the output token. The saddle-point equations,
\begin{equation}
    \frac{\partial G_S^{\mathrm{spec}}}{\partial q_m}
    +\alpha
    \frac{\partial G_E^{\mathrm{spec}}}{\partial q_m}
    =0,
    \qquad m=1,\ldots,\mathrm{r}_\Sigma,
    \label{eq:main_dict_spectral_sp}
\end{equation}
balance the entropic cost of constraining the student against the reduction in label uncertainty. At the saddle $\bm q^\star$, the Bayes predictor selects the token with largest posterior probability, giving
\begin{equation}
    \epsilon_{\mathrm{Bayes}}(\mathbf  Z,\alpha)
    =\mathbb E\left[1-\max_k p_k\right]
    \label{eq:main_dict_spectral_error}
\end{equation}
In Fig.~\ref{fig:iso_vs_aniso} we compare the values of generalization error obtained by solving saddle point equations for Bayes-optimal predictor, with the simulated lowest generalization error achieved by numerical network students trained using generalized cross-entropy loss, Eq. \eqref{eq:loss_dict}.  

\subsection{Scaling of learning curves}
\label{sec:scaling_curves}
To compare learning curves across dimensions, we introduce the rescaled load
\begin{equation}
    \widetilde\alpha
    =\frac{\alpha}{\mathrm{r}_Z^2}
    =\frac{P}{N\mathrm{r}_Z^2},
\end{equation}
which normalizes the load by the rank-dependent entropic scale of the spectral closure, as derived in Appendix~\ref{app:vocabulary_replica}. The saddle-point equations then read
\begin{equation}
    \frac{q_m}{1-q_m}
    =-2\mathrm{r}_Z\widetilde\alpha
    \frac{\partial H_Z}{\partial q_m}
    \label{eq:sp_with_alpha_tilde}
\end{equation}
For a generic vocabulary, $\mathrm{r}_Z=\min(d,D)$, then when $d\leq D$ the vocabulary spans the full embedding space and $\mathrm{r}_Z=d$, while if $d>D$ its span has dimension $\mathrm{r}_Z=D$. The rank-dependent normalization therefore saturates once $d$ exceeds $D$.

\begin{figure}
    \centering
    \includegraphics[width=0.45\linewidth]{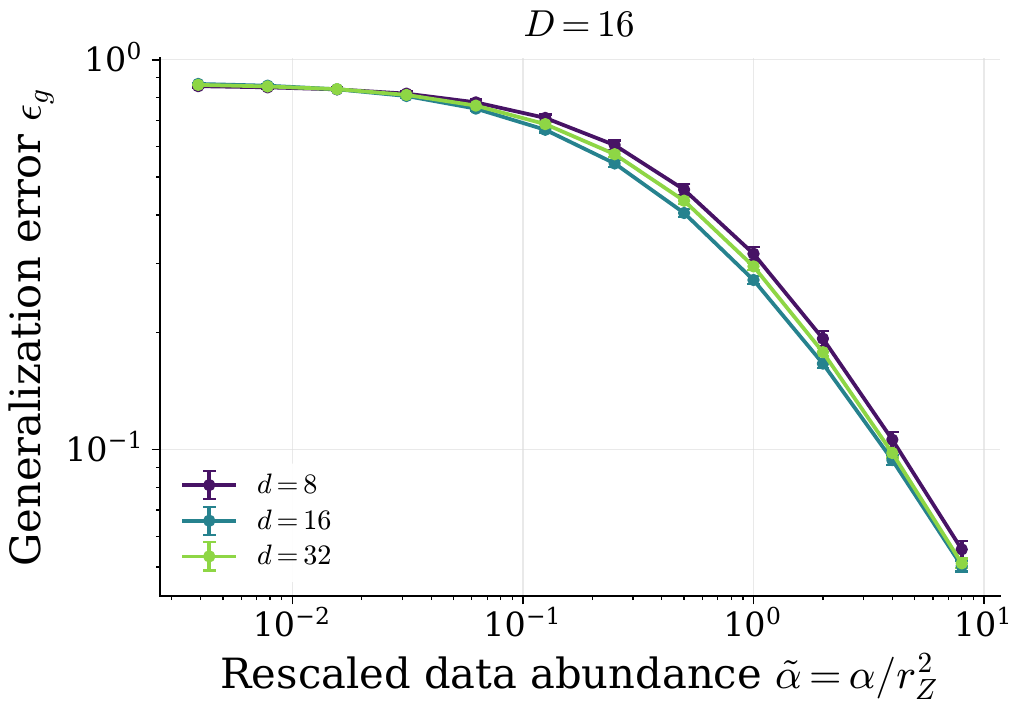}
    \includegraphics[width=0.45\linewidth]{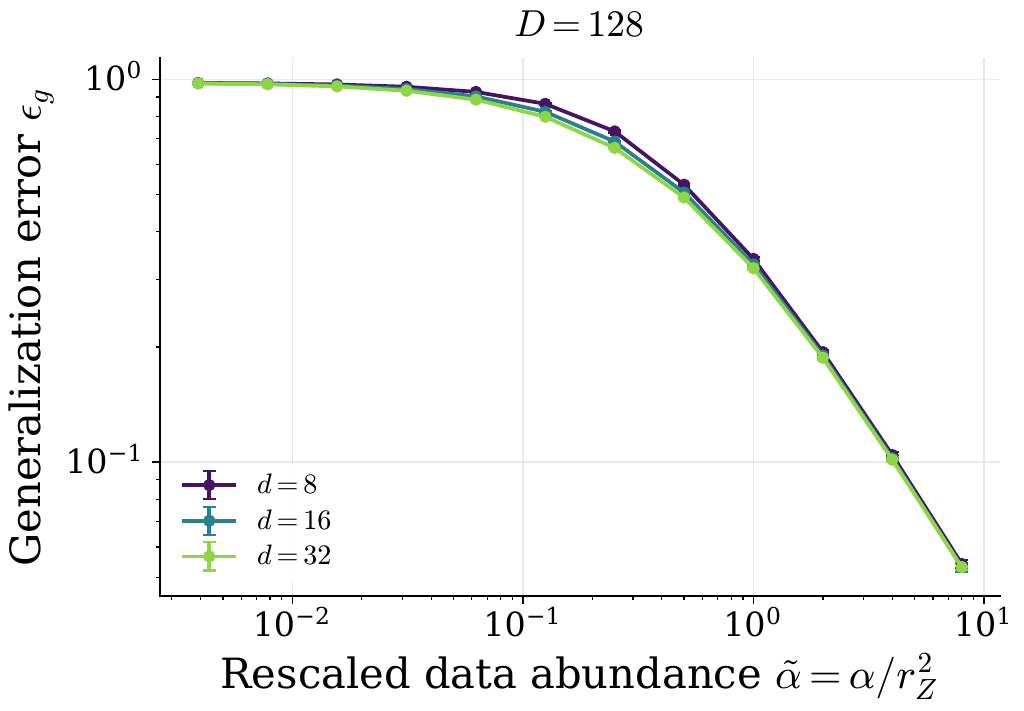}
    \caption{\textbf{Learning curves scale correctly by normalizing amount of data via} $\widetilde\alpha$. (Left) D=16 case; we note that here $d=32>D$ and in that case $\mathrm{r}_Z$ saturates and the curves do not shift anymore when plotted against $\alpha$, as occurs in Fig.~\ref{fig:iso_vs_aniso}. (Right) D=128; in this case all curves have $d<D$ and therefore the problem remains full rank. We observe that, even if the curves are rescaled correctly, there is still a dependence on $d,D$ due to the term $H_Z$ that depends explicitly on the vocabulary.}
    \label{fig:alpha_tilde}
\end{figure}

In the spectral closure, constraining each mode has an entropic cost proportional to $\mathrm{r}_Z$, and there are $\mathrm{r}_\Sigma$ modes to learn. The total entropic scale is therefore $\mathrm{r}_Z\mathrm{r}_\Sigma$, of order $\mathrm{r}_Z^2$, while the data contribution scales with $\alpha=P/N$. Increasing the rank thus requires more examples to maintain comparable data abundance relative to this learning cost, and then we expect from the solution of Eq. \eqref{eq:sp_with_alpha_tilde} a trivial rescaling of learning curves as $\widetilde\alpha=\alpha/\mathrm{r}_Z^2$, as observed in Fig.~\ref{fig:alpha_tilde}. The conditional entropy $H_Z$ still depends on the vocabulary geometry, including its spectrum and decision cells. Differences between curves at fixed $\widetilde\alpha$ therefore reveal dimension dependence beyond the explicit rank normalization.

\subsection{The role of vocabulary embeddings geometry}
\label{sec:spec_results}
\begin{figure}[ht!]
    \centering
    \includegraphics[width=0.48\linewidth]{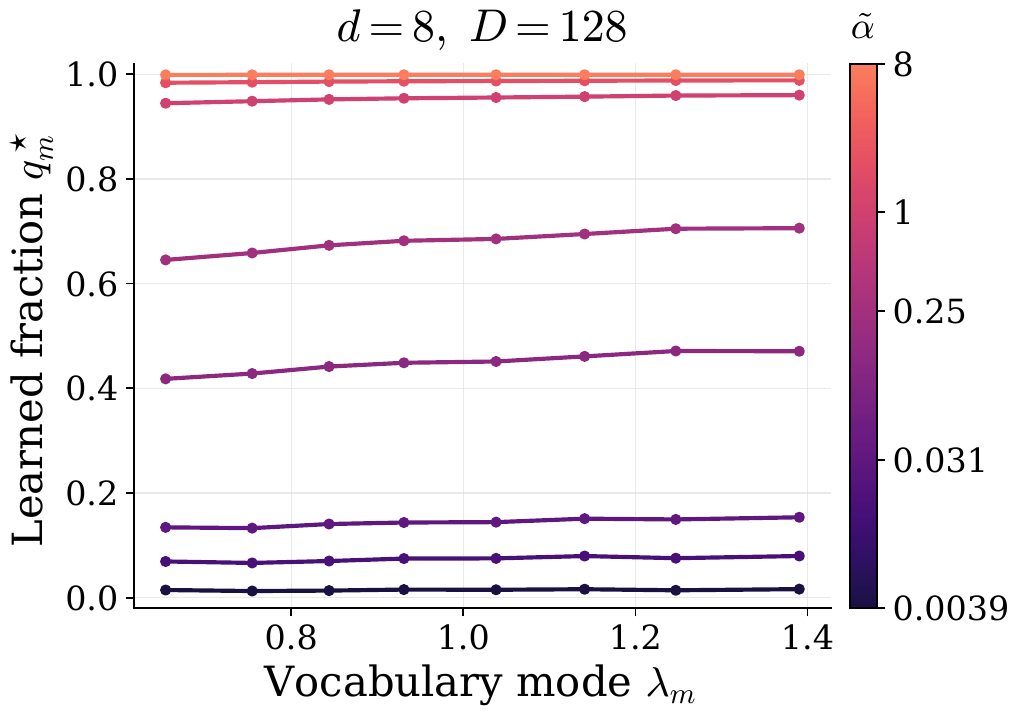}
    \includegraphics[width=0.48\linewidth]{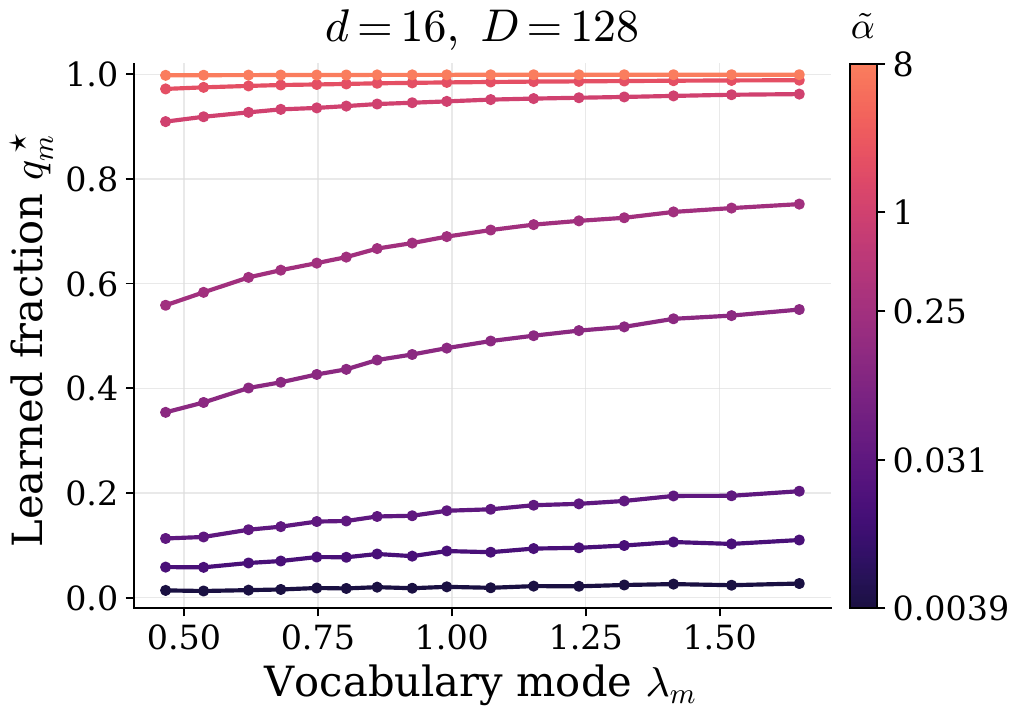}
    \includegraphics[width=0.48\linewidth]{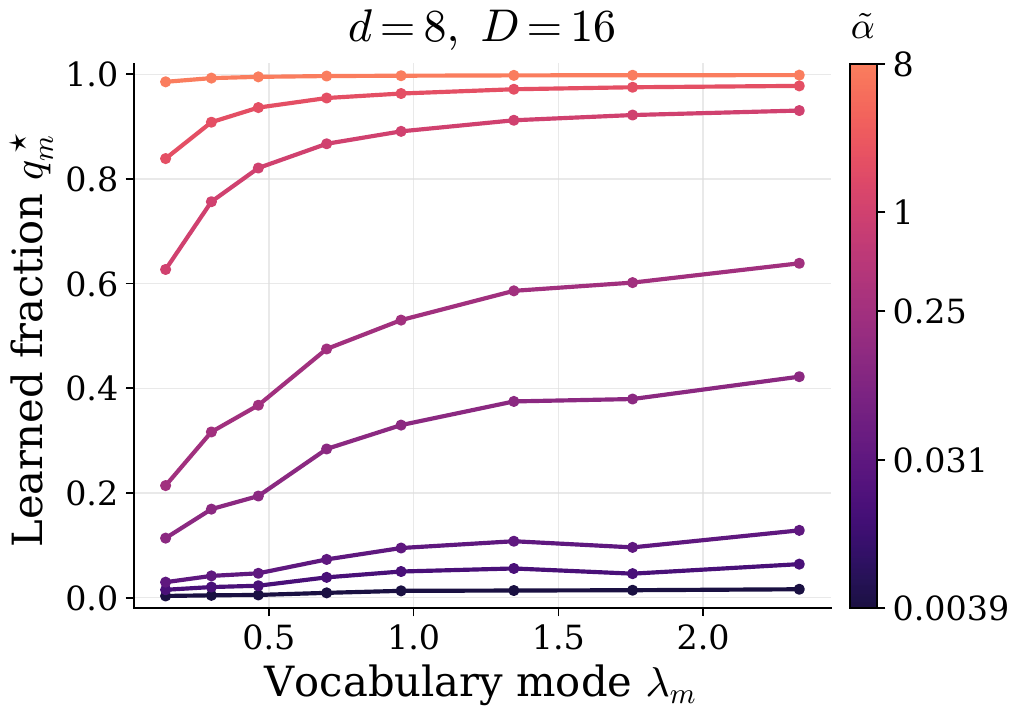}
    \includegraphics[width=0.48\linewidth]{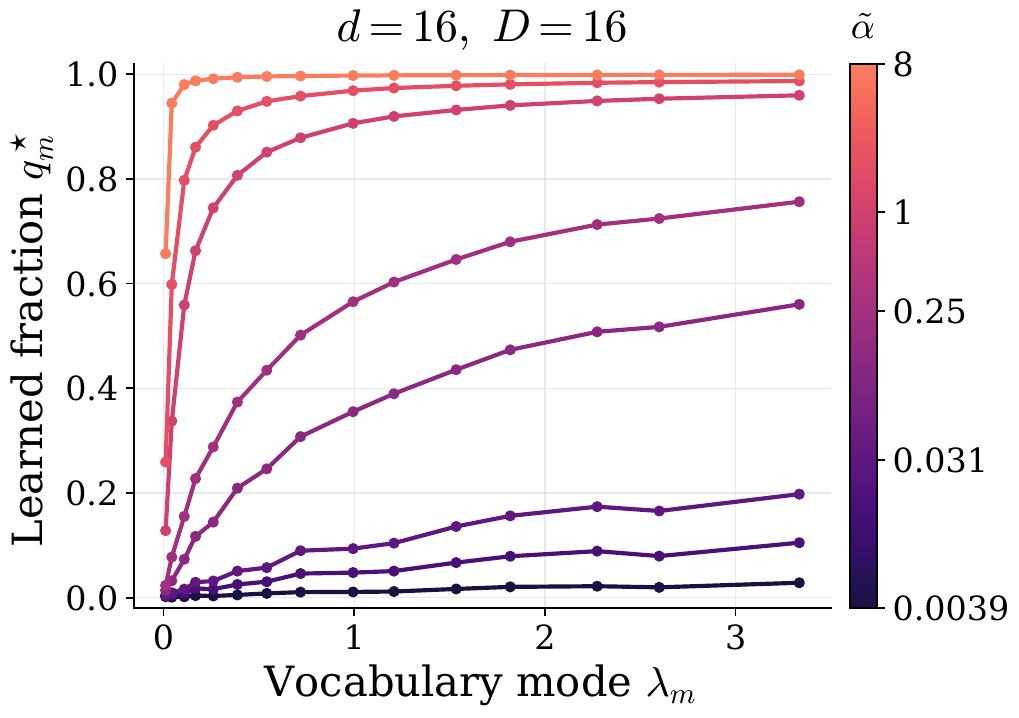}
    \caption{\textbf{The optimal strategy is to learn high variance vocabulary embeddings direction first, then weaker directions as more data is available.} (top left) case $d=8, D=128$, (top right) case $d=16$, $D=128$, (bottom left) case $d=8$, $D=16$, (bottom right) case $d=16$, $D=16$. In all cases we observe that the normalized overlap $q_m^*$ increase monotonically with the corresponding $\lambda_m$. Therefore, even if the task is isotropic ({learning an isotropic spherical teacher}), the anisotropy of the vocabulary affects the geometry of the inference. Saddle-point from replica computation defines implicitly the relations $q^*_m(\lambda_m)$ that control how learning happens in each spectral direction $m$.}
    \label{fig:q_vs_l_D6}
\end{figure}

The vocabulary eigenvalues characterize the geometrical properties of the problem, but do not alone determine which directions are most useful for prediction: their relevance is measured by $-\partial H_Z/\partial q_m$, meaning that a direction is informative when learning it resolves uncertainty between competing tokens. Consequently modes with higher vocabulary eigenvalues $\lambda_m$ are generically expected to produce higher $q_m^*$, but in principle even a direction with small vocabulary variance may be important for classification if it happens to discriminate competing tokens. To investigate quantitatively how vocabulary covariance spectrum affects learning in our random vocabulary setting, we study the distribution of spectral order parameters $\{q^*_m\}$ from the solution of saddle-point, Eq. \eqref{eq:main_dict_spectral_sp}. Fig.~\ref{fig:q_vs_l_D6} shows that indeed vocabulary directions with higher eigenvalues $\lambda_m$ also have higher overlap $q^*_m$: learning stronger modes of the vocabulary first is the optimal strategy when data is scarce. As data abundance $\widetilde\alpha$ increases, the high-variance directions overlap $q^*_m$ saturates to 1, but progressively the model learns also the weaker directions of the vocabulary. We observe that this effect is stronger for more anisotropic vocabularies, and this property depends on $d,D$ for embeddings extracted from an isotropic distribution, because anisotropy is due only to finiteness of vocabulary elements. In particular for $d\ll D$ all modes are learned almost equally, while as $d$ gets closer to $D$ the minimum value of $\Lambda_m$ decreases and this filtering effect becomes more evident.

\section{Discussion}
Inspired by LLMs models we promote to multidimensional variables classical perceptron theory introducing two different generalizations, the continuous embedding and the vocabulary embedding cases. We define a simplified framework in order to apply statistical mechanics to this new setting, showing the appearance of new phenomena that have not a counterpart for neural networks with one-dimensional variables.  \\
\\In the case of spherical embeddings, we show that increasing the embedding dimension $d$ qualitatively changes the learning problem: for $d>1$, the teacher can be exactly recovered from a finite number of examples per input position, with a critical load $\alpha_{\rm rec}=d^2/(d-1)$. We show that spherical cross-entropy reaches this exact-recovery regime in the large-$\lambda$ limit, whereas for $d=1$ the corresponding learning rule approaches the maximum-margin classifier, whose generalization error remains finite at any finite load \cite{soudry18}. Then, we study the ability of the network to store random associations between input sequences and output labels, that can be defined constraining the projection of the output into the label, that coincides at $d=1$ with Gardner's storage, and a new definition involving the angular alignment, that is defined only for $d>1$. In both cases we show that for margins $\kappa=\mathcal{O}(1/\sqrt{d})$, the capacity increases proportionally to $d^2$, similarly to factual recall \cite{nichaniICLR2025,giorlandino2026factual} and more than other associative memories in a close setting \cite{nicoletti2025statistical, barney2026quantum}. \\
\\For vocabulary classification, the novelty with respect to previous teacher-student literature is that generalization is no longer controlled only by the alignment with a continuous teacher direction: the prediction is a competition between $D$ cells whose geometry is fixed by the vocabulary. Learning becomes a competition between \emph{spectral variance} and \emph{task relevance}: the posterior uncertainty can shrink at different rates along different embedding modes depending on how strongly those modes help to distinguish the vocabulary embeddings. In the case of spherical random vocabulary embeddings results suggest that the optimal way to solve the inference is to learn first the task directions that are "more visible", i.e. in which the vocabulary has more variance, and then to use "less visible" directions as there are more data, even if the inference task itself is isotropic. We note that an implicit bias towards learning high-variance modes first is also observed in other contexts of machine learning, such as in the generative and training dynamics of diffusion models \cite{achilli2026losingdimensionsgeometricmemorization, wang2026analyticaltheoryspectralbias, nicoletti2026interplaydatastructureimbalance}.

As final remarks, we considered a tractable model in order to study the role of embeddings in the learning process, with several differences with real settings of LLMs. Mainly, we have an explicit positional dependency of the neural models, compared to positional encoding; in real settings learning of the embedding is done jointly with learning of the model; and the task is supervised token classification, missing the role of the sequence temporal structure, necessary to model how LLMs learn to predict next token in a sequence. 

\section{Acknowledgments}
We thank Brandon Livio Amnesi, Sebastian Goldt and Daniel J. Korchinski for insightful discussions. FD acknowledges funding from the Bando Ricerca Scientifica 2025 - Avvio alla Ricerca (D. R. 2155/2025) of Sapienza Università di Roma, project B83C25004300005 - VEctor-perceptron for Self-attention: Theory and Assessment (VESTA). This study was conducted using the DARIAH HPC-AI cluster at CNR-NANOTEC in Lecce, funded by the "MUR PON Ricerca e Innovazione 2014-2020" project, code PIR01\_00022 and H2IOSC Project - Humanities and cultural Heritage Italian Open Science Cloud funded by the European Union – NextGenerationEU – NRRP M4C2 - Project code IR0000029. We used generative AI models to assist with literature review, deriving and checking analytical expressions, writing the simulation code, and drafting and revising the text. All results were verified by the authors, who take full responsibility for the content of this work.


\bibliography{refs}

@article{krogh1992generalization,
  title   = {Generalization in a linear perceptron in the presence of noise},
  author  = {Krogh, Anders and Hertz, John A.},
  journal = {Journal of Physics A: Mathematical and General},
  volume  = {25},
  number  = {5},
  pages   = {1135},
  year    = {1992},
  doi     = {10.1088/0305-4470/25/5/020},
  url     = {https://doi.org/10.1088/0305-4470/25/5/020}
}

@article{watkin1993statistical,
  title   = {The statistical mechanics of learning a rule},
  author  = {Watkin, Timothy L. H. and Rau, Albrecht and Biehl, Michael},
  journal = {Rev. Mod. Phys.},
  volume  = {65},
  issue   = {2},
  pages   = {499--556},
  year    = {1993},
  publisher = {American Physical Society},
  doi     = {10.1103/RevModPhys.65.499},
  url     = {https://link.aps.org/doi/10.1103/RevModPhys.65.499}
}

@inproceedings{press2017output,
  title     = {Using the Output Embedding to Improve Language Models},
  author    = {Press, Ofir and Wolf, Lior},
  booktitle = {Proceedings of the 15th Conference of the European Chapter of the Association for Computational Linguistics: Volume 2, Short Papers},
  pages     = {157--163},
  year      = {2017},
  publisher = {Association for Computational Linguistics},
  url       = {https://aclanthology.org/E17-2025}
}

@inproceedings{NEURIPS2020_1457c0d6,
 author = {Brown, Tom and Mann, Benjamin and Ryder, Nick and Subbiah, Melanie and Kaplan, Jared D and Dhariwal, Prafulla and Neelakantan, Arvind and Shyam, Pranav and Sastry, Girish and Askell, Amanda and Agarwal, Sandhini and Herbert-Voss, Ariel and Krueger, Gretchen and Henighan, Tom and Child, Rewon and Ramesh, Aditya and Ziegler, Daniel and Wu, Jeffrey and Winter, Clemens and Hesse, Chris and Chen, Mark and Sigler, Eric and Litwin, Mateusz and Gray, Scott and Chess, Benjamin and Clark, Jack and Berner, Christopher and McCandlish, Sam and Radford, Alec and Sutskever, Ilya and Amodei, Dario},
 booktitle = {Advances in Neural Information Processing Systems},
 editor = {H. Larochelle and M. Ranzato and R. Hadsell and M.F. Balcan and H. Lin},
 pages = {1877--1901},
 publisher = {Curran Associates, Inc.},
 title = {Language Models are Few-Shot Learners},
 url = {https://proceedings.neurips.cc/paper_files/paper/2020/file/1457c0d6bfcb4967418bfb8ac142f64a-Paper.pdf},
 volume = {33},
 year = {2020}
}

@misc{radford2018improving,
  author = {Radford, Alec and Narasimhan, Karthik and Salimans, Tim and Sutskever, Ilya},
  title = {Improving language understanding by generative pre-training},
  howpublished = {OpenAI technical report},
  year = {2018},
  url = {https://cdn.openai.com/research-covers/language-unsupervised/language_understanding_paper.pdf}
}

@inproceedings{NIPS2017_3f5ee243,
 author = {Vaswani, Ashish and Shazeer, Noam and Parmar, Niki and Uszkoreit, Jakob and Jones, Llion and Gomez, Aidan N and Kaiser, \L ukasz and Polosukhin, Illia},
 booktitle = {Advances in Neural Information Processing Systems},
 editor = {I. Guyon and U. Von Luxburg and S. Bengio and H. Wallach and R. Fergus and S. Vishwanathan and R. Garnett},
 pages = {},
 publisher = {Curran Associates, Inc.},
 title = {Attention is All you Need},
 url = {https://proceedings.neurips.cc/paper_files/paper/2017/file/3f5ee243547dee91fbd053c1c4a845aa-Paper.pdf},
 volume = {30},
 year = {2017}
}

@inproceedings{NEURIPS2022_b1efde53,
 author = {Ouyang, Long and Wu, Jeffrey and Jiang, Xu and Almeida, Diogo and Wainwright, Carroll and Mishkin, Pamela and Zhang, Chong and Agarwal, Sandhini and Slama, Katarina and Ray, Alex and Schulman, John and Hilton, Jacob and Kelton, Fraser and Miller, Luke and Simens, Maddie and Askell, Amanda and Welinder, Peter and Christiano, Paul F and Leike, Jan and Lowe, Ryan},
 booktitle = {Advances in Neural Information Processing Systems},
 doi = {10.52202/068431-2011},
 editor = {S. Koyejo and S. Mohamed and A. Agarwal and D. Belgrave and K. Cho and A. Oh},
 pages = {27730--27744},
 publisher = {Curran Associates, Inc.},
 title = {Training language models to follow instructions with human feedback},
 url = {https://proceedings.neurips.cc/paper_files/paper/2022/file/b1efde53be364a73914f58805a001731-Paper-Conference.pdf},
 volume = {35},
 year = {2022}
}

@inproceedings{sompo_mignacco,
 author = {Tiberi, Lorenzo and Mignacco, Francesca and Irie, Kazuki and Sompolinsky, Haim},
 booktitle = {Advances in Neural Information Processing Systems},
 doi = {10.52202/079017-2315},
 editor = {A. Globerson and L. Mackey and D. Belgrave and A. Fan and U. Paquet and J. Tomczak and C. Zhang},
 pages = {72710--72753},
 publisher = {Curran Associates, Inc.},
 title = {Dissecting the Interplay of Attention Paths in a Statistical Mechanics Theory of Transformers},
 url = {https://proceedings.neurips.cc/paper_files/paper/2024/file/8523a98265ceae12afd34113aa6c5cca-Paper-Conference.pdf},
 volume = {37},
 year = {2024}
}

@article{maillard,
  title = {Bilinear Sequence Regression: A Model for Learning from Long Sequences of High-Dimensional Tokens},
  author = {Erba, Vittorio and Troiani, Emanuele and Biggio, Luca and Maillard, Antoine and Zdeborov{\'a}, Lenka},
  journal = {Phys. Rev. X},
  volume = {15},
  issue = {2},
  pages = {021092},
  numpages = {43},
  year = {2025},
  month = {Jun},
  publisher = {American Physical Society},
  doi = {10.1103/l4p2-vrxt},
  url = {https://link.aps.org/doi/10.1103/l4p2-vrxt}
}

@inproceedings{Fab_attention_indexed,
 author = {Boncoraglio, Fabrizio and Troiani, Emanuele and Erba, Vittorio and Zdeborov\'{a}, Lenka},
 booktitle = {Advances in Neural Information Processing Systems},
 doi = {10.52202/085713-3509},
 editor = {D. Belgrave and C. Zhang and H. Lin and R. Pascanu and P. Koniusz and M. Ghassemi and N. Chen},
 pages = {105029--105074},
 publisher = {Curran Associates, Inc.},
 title = {Bayes optimal learning of attention-indexed models},
 url = {https://proceedings.neurips.cc/paper_files/paper/2025/file/9740e3572538a1a9132d8d4ad2e8bf72-Paper-Conference.pdf},
 volume = {38, Main Conference},
 year = {2025}
}

@article{E_Gardner_1988,
doi = {10.1088/0305-4470/21/1/030},
url = {https://doi.org/10.1088/0305-4470/21/1/030},
year = {1988},
month = {jan},
publisher = {},
volume = {21},
number = {1},
pages = {257},
author = {E Gardner},
title = {The space of interactions in neural network models},
journal = {Journal of Physics A: Mathematical and General}
}

@article{soudry18,
  author  = {Daniel Soudry and Elad Hoffer and Mor Shpigel Nacson and Suriya Gunasekar and Nathan Srebro},
  title   = {The Implicit Bias of Gradient Descent on Separable Data},
  journal = {Journal of Machine Learning Research},
  year    = {2018},
  volume  = {19},
  number  = {70},
  pages   = {1--57},
  url     = {http://jmlr.org/papers/v19/18-188.html}
}

@misc{RMS,
      title={Root Mean Square Layer Normalization}, 
      author={Biao Zhang and Rico Sennrich},
      year={2019},
      eprint={1910.07467},
      archivePrefix={arXiv},
      primaryClass={cs.LG},
      url={https://arxiv.org/abs/1910.07467}, 
}

@inproceedings{bordelon_ict,
 author = {Bordelon, Blake and Letey, Mary and Pehlevan, Cengiz},
 booktitle = {International Conference on Learning Representations},
 editor = {C. Vondrick and B. Hariharan and C. Raffel and L. Pinto and D. Yang and A. Faust},
 pages = {21749--21779},
 title = {Theory of Scaling Laws for In-Context Regression: Depth, Width, Context and Time},
 url = {https://proceedings.iclr.cc/paper_files/paper/2026/file/24d36eee157559e0d2549455fba28f6a-Paper-Conference.pdf},
 volume = {2026},
 year = {2026}
}

@misc{barnfield2026factual,
      title={Sharp Capacity Thresholds in Linear Associative Memory: From Top-1 Retrieval to Tail-Average Learning}, 
      author={Nicholas Barnfield and Juno Kim and Eshaan Nichani and Jason D. Lee and Yue M. Lu},
      year={2026},
      eprint={2605.05189},
      archivePrefix={arXiv},
      primaryClass={stat.ML},
      url={https://arxiv.org/abs/2605.05189}, 
}

@misc{giorlandino2026factual,
      title={Factual recall in linear associative memories: sharp asymptotics and mechanistic insights}, 
      author={Alessio Giorlandino and Sebastian Goldt and Antoine Maillard},
      year={2026},
      eprint={2605.10795},
      archivePrefix={arXiv},
      primaryClass={stat.ML},
      url={https://arxiv.org/abs/2605.10795}, 
}

@article{sompo_perc,
  title = {Learning from examples in large neural networks},
  author = {Sompolinsky, H. and Tishby, N. and Seung, H. S.},
  journal = {Phys. Rev. Lett.},
  volume = {65},
  issue = {13},
  pages = {1683--1686},
  numpages = {0},
  year = {1990},
  month = {Sep},
  publisher = {American Physical Society},
  doi = {10.1103/PhysRevLett.65.1683},
  url = {https://link.aps.org/doi/10.1103/PhysRevLett.65.1683}
}

@inproceedings{Engel2001StatisticalMO,
  title={Statistical Mechanics of Learning},
  author={Andreas Engel and Christian Van den Broeck},
  year={2001},
  url={https://api.semanticscholar.org/CorpusID:60723885}
}

@inproceedings{
dosovitskiy2021an,
title={An Image is Worth 16x16 Words: Transformers for Image Recognition at Scale},
author={Alexey Dosovitskiy and Lucas Beyer and Alexander Kolesnikov and Dirk Weissenborn and Xiaohua Zhai and Thomas Unterthiner and Mostafa Dehghani and Matthias Minderer and Georg Heigold and Sylvain Gelly and Jakob Uszkoreit and Neil Houlsby},
booktitle={International Conference on Learning Representations},
year={2021},
url={https://openreview.net/forum?id=YicbFdNTTy}
}

@misc{sennrich2016token,
      title={Neural Machine Translation of Rare Words with Subword Units}, 
      author={Rico Sennrich and Barry Haddow and Alexandra Birch},
      year={2016},
      eprint={1508.07909},
      archivePrefix={arXiv},
      primaryClass={cs.CL},
      url={https://arxiv.org/abs/1508.07909}, 
}

@misc{kudo2018token,
      title={SentencePiece: A simple and language independent subword tokenizer and detokenizer for Neural Text Processing}, 
      author={Taku Kudo and John Richardson},
      year={2018},
      eprint={1808.06226},
      archivePrefix={arXiv},
      primaryClass={cs.CL},
      url={https://arxiv.org/abs/1808.06226}, 
}

@inproceedings{Daniel_1,
 author = {Korchinski, Daniel and Karkada, Dhruva and Bahri, Yasaman and Wyart, Matthieu},
 booktitle = {Advances in Neural Information Processing Systems},
 doi = {10.52202/085713-2075},
 editor = {D. Belgrave and C. Zhang and H. Lin and R. Pascanu and P. Koniusz and M. Ghassemi and N. Chen},
 pages = {62069--62097},
 publisher = {Curran Associates, Inc.},
 title = {On the Emergence of Linear Analogies in Word Embeddings},
 url = {https://proceedings.neurips.cc/paper_files/paper/2025/file/59997f550daa5880f50939927181177c-Paper-Conference.pdf},
 volume = {38, Main Conference},
 year = {2025}
}

@misc{Daniel_2,
      title={Symmetry in language statistics shapes the geometry of model representations}, 
      author={Dhruva Karkada and Daniel J. Korchinski and Andres Nava and Matthieu Wyart and Yasaman Bahri},
      year={2026},
      eprint={2602.15029},
      archivePrefix={arXiv},
      primaryClass={cs.LG},
      url={https://arxiv.org/abs/2602.15029}, 
}

@article{
Lu_ICT,
author = {Yue M. Lu  and Mary Letey  and Jacob A. Zavatone-Veth  and Anindita Maiti  and Cengiz Pehlevan },
title = {Asymptotic theory of in-context learning by linear attention},
journal = {Proceedings of the National Academy of Sciences},
volume = {122},
number = {28},
pages = {e2502599122},
year = {2025},
doi = {10.1073/pnas.2502599122},
URL = {https://www.pnas.org/doi/abs/10.1073/pnas.2502599122},
eprint = {https://www.pnas.org/doi/pdf/10.1073/pnas.2502599122}}

@misc{letey2026sequential,
      title={Sequential Correlations Change In-Context Learning: Effective Context Length and Architectural Mismatch}, 
      author={Mary Letey and Yue M. Lu and Cengiz Pehlevan and Jacob Zavatone-Veth},
      year={2026},
      eprint={2607.03660},
      archivePrefix={arXiv},
      primaryClass={stat.ML},
      url={https://arxiv.org/abs/2607.03660}, 
}

@misc{scott2021vMF,
      title={von Mises-Fisher Loss: An Exploration of Embedding Geometries for Supervised Learning}, 
      author={Tyler R. Scott and Andrew C. Gallagher and Michael C. Mozer},
      year={2021},
      eprint={2103.15718},
      archivePrefix={arXiv},
      primaryClass={cs.LG},
      url={https://arxiv.org/abs/2103.15718}, 
}

@article{besag1974spatial,
  title={Spatial interaction and the statistical analysis of lattice systems},
  author={Besag, Julian},
  journal={Journal of the Royal Statistical Society: Series B (Methodological)},
  volume={36},
  number={2},
  pages={192--225},
  year={1974},
  publisher={Wiley Online Library}
}

@article{PL_potts,
  title = {Improved contact prediction in proteins: Using pseudolikelihoods to infer Potts models},
  author = {Ekeberg, Magnus and L{\"o}vkvist, Cecilia and Lan, Yueheng and Weigt, Martin and Aurell, Erik},
  journal = {Phys. Rev. E},
  volume = {87},
  issue = {1},
  pages = {012707},
  numpages = {16},
  year = {2013},
  month = {Jan},
  publisher = {American Physical Society},
  doi = {10.1103/PhysRevE.87.012707},
  url = {https://link.aps.org/doi/10.1103/PhysRevE.87.012707}
}

@article{DAMICO2026131497,
title = {Pseudo-likelihood produces associative memories able to generalize, even for asymmetric couplings},
journal = {Physica A: Statistical Mechanics and its Applications},
volume = {692},
pages = {131497},
year = {2026},
issn = {0378-4371},
doi = {https://doi.org/10.1016/j.physa.2026.131497},
url = {https://www.sciencedirect.com/science/article/pii/S0378437126002335},
author = {Francesco D’Amico and Dario Bocchi and Luca Maria {Del Bono} and Saverio Rossi and Matteo Negri}
}

@article{nicoletti2025statistical,
  title={Statistical mechanics of vector Hopfield network near and above saturation},
  author={Nicoletti, Flavio and D’Amico, Francesco and Negri, Matteo},
  journal={Journal of Physics A: Mathematical and Theoretical},
  volume={58},
  number={50},
  pages={505005},
  year={2025},
  publisher={IOP Publishing},
  url={https://iopscience.iop.org/article/10.1088/1751-8121/ae2bd0/meta},
  doi={10.1088/1751-8121/ae2bd0/meta}
}

@article{barney2026quantum,
  title={Quantum-stabilized patterns in a vector Hopfield network},
  author={Barney, Richard D and Bhattacharjee, Sharba and Galitski, Victor and Agarwal, Kartiek and Martin, Ivar},
  journal={arXiv preprint arXiv:2606.06597},
  year={2026},
  url={https://arxiv.org/abs/2606.06597},
  doi={2606.06597}
}

@misc{galitski2026highcapacitygeneralizedhopfieldnetworks,
      title={High-Capacity Generalized Hopfield Networks}, 
      author={Victor Galitski},
      year={2026},
      eprint={2608.08226},
      archivePrefix={arXiv},
      primaryClass={cond-mat.stat-mech},
      url={https://arxiv.org/abs/2608.08226},
      doi={2608.08226}
}

@inproceedings{nichaniICLR2025,
 author = {Nichani, Eshaan and Lee, Jason and Bietti, Alberto},
 booktitle = {International Conference on Learning Representations},
 editor = {Y. Yue and A. Garg and N. Peng and F. Sha and R. Yu},
 pages = {4166--4207},
 title = {Understanding Factual Recall in Transformers via Associative Memories},
 url = {https://proceedings.iclr.cc/paper_files/paper/2025/file/0bf9f909c24c8879d1b7f86fa50a9e49-Paper-Conference.pdf},
 volume = {2025},
 year = {2025}
}

@misc{gao2026cluster,
      title={Clustered Attractor Manifolds and Dynamical Condensation in Self-Attention}, 
      author={Qucheng Gao and Zuyi Yang and Xiao Chen},
      year={2026},
      eprint={2608.08922},
      archivePrefix={arXiv},
      primaryClass={cond-mat.dis-nn},
      url={https://arxiv.org/abs/2608.08922}, 
}

@misc{gao2026repulsive,
      title={Nonequilibrium Phases of Repulsive Self-Attention: Chaos, Attention Condensation, and Emergent Locality}, 
      author={Qucheng Gao and Zuyi Yang and Xiao Chen},
      year={2026},
      eprint={2609.28448},
      archivePrefix={arXiv},
      primaryClass={cond-mat.dis-nn},
      url={https://arxiv.org/abs/2609.28448}, 
}

@article{opper1991,
  title = {Generalization performance of Bayes optimal classification algorithm for learning a perceptron},
  author = {Opper, Manfred and Haussler, David},
  journal = {Phys. Rev. Lett.},
  volume = {66},
  issue = {20},
  pages = {2677--2680},
  numpages = {0},
  year = {1991},
  month = {May},
  publisher = {American Physical Society},
  doi = {10.1103/PhysRevLett.66.2677},
  url = {https://link.aps.org/doi/10.1103/PhysRevLett.66.2677}
}

@inproceedings{damico_ICLR,
 author = {D'Amico, Francesco and Bocchi, Dario and Negri, Matteo},
 booktitle = {International Conference on Learning Representations},
 editor = {C. Vondrick and B. Hariharan and C. Raffel and L. Pinto and D. Yang and A. Faust},
 pages = {54318--54344},
 title = {Implicit bias produces neural scaling laws in learning curves, from perceptrons to deep networks},
 url = {https://proceedings.iclr.cc/paper_files/paper/2026/file/5934ef82865c4456fc708c311801169d-Paper-Conference.pdf},
 volume = {2026},
 year = {2026}
}

@inproceedings{aubin_convex_opt,
 author = {Aubin, Benjamin and Krzakala, Florent and Lu, Yue and Zdeborov\'{a}, Lenka},
 booktitle = {Advances in Neural Information Processing Systems},
 editor = {H. Larochelle and M. Ranzato and R. Hadsell and M.F. Balcan and H. Lin},
 pages = {12199--12210},
 publisher = {Curran Associates, Inc.},
 title = {Generalization error in high-dimensional perceptrons: Approaching Bayes error with convex optimization},
 url = {https://proceedings.neurips.cc/paper_files/paper/2020/file/8f4576ad85410442a74ee3a7683757b3-Paper.pdf},
 volume = {33},
 year = {2020}
}

@misc{wang2026analyticaltheoryspectralbias,
      title={An Analytical Theory of Spectral Bias in the Learning Dynamics of Diffusion Models}, 
      author={Binxu Wang and Cengiz Pehlevan},
      year={2026},
      eprint={2503.03206},
      archivePrefix={arXiv},
      primaryClass={cs.LG},
      url={https://arxiv.org/abs/2503.03206},
      doi={2503.03206}
}

@misc{nicoletti2026interplaydatastructureimbalance,
      title={The Interplay of Data Structure and Imbalance in the Learning Dynamics of Diffusion Models}, 
      author={Flavio Nicoletti and Chenxiao Ma and Enrico Ventura and Luca Saglietti and Stefano Sarao Mannelli},
      year={2026},
      eprint={2605.06367},
      archivePrefix={arXiv},
      primaryClass={stat.ML},
      url={https://arxiv.org/abs/2605.06367}, 
      doi={2605.06367}
}

@misc{achilli2026losingdimensionsgeometricmemorization,
      title={Losing dimensions: Geometric memorization in generative diffusion}, 
      author={Beatrice Achilli and Enrico Ventura and Gianluigi Silvestri and Bao Pham and Gabriel Raya and Dmitry Krotov and Carlo Lucibello and Luca Ambrogioni},
      year={2026},
      eprint={2410.08727},
      archivePrefix={arXiv},
      primaryClass={stat.ML},
      url={https://arxiv.org/abs/2410.08727},
      doi={2410.08727}
}

@article{gardner1988optimalstorage,
doi = {10.1088/0305-4470/21/1/031},
url = {https://doi.org/10.1088/0305-4470/21/1/031},
year = {1988},
month = {jan},
publisher = {},
volume = {21},
number = {1},
pages = {271},
author = {E Gardner and B Derrida},
title = {Optimal storage properties of neural network models},
journal = {Journal of Physics A: Mathematical and General}
}

@inproceedings{cagnetta2024structure,
 author = {Cagnetta, Francesco and Wyart, Matthieu},
 booktitle = {Advances in Neural Information Processing Systems},
 doi = {10.52202/079017-2645},
 editor = {A. Globerson and L. Mackey and D. Belgrave and A. Fan and U. Paquet and J. Tomczak and C. Zhang},
 pages = {83119--83163},
 publisher = {Curran Associates, Inc.},
 title = {Towards a theory of how the structure of language is acquired by deep neural networks},
 url = {https://proceedings.neurips.cc/paper_files/paper/2024/file/9740da1c07c7b451af14e11523f95271-Paper-Conference.pdf},
 volume = {37},
 year = {2024}
}

@article{hertz1989phase,
  title={Phase transitions in simple learning},
  author={Hertz, JA and Krogh, A and Thorbergsson, GI},
  journal={Journal of Physics A: Mathematical and General},
  volume={22},
  number={12},
  pages={2133--2150},
  year={1989}
}

@book{muirhead2009aspects,
  title={Aspects of multivariate statistical theory},
  author={Muirhead, Robb J},
  year={2009},
  publisher={John Wiley \& Sons}
}

@book{mezard1987spin,
  title={Spin glass theory and beyond: An Introduction to the Replica Method and Its Applications},
  author={M{\'e}zard, Marc and Parisi, Giorgio and Virasoro, Miguel Angel},
  volume={9},
  year={1987},
  publisher={World Scientific Publishing Company}
}

@article{clark2026transient,
  title = {Transient dynamics of associative memory models},
  author = {Clark, David G.},
  journal = {Phys. Rev. E},
  volume = {113},
  issue = {5},
  pages = {054301},
  numpages = {19},
  year = {2026},
  month = {May},
  publisher = {American Physical Society},
  doi = {10.1103/42y2-bsh1},
  url = {https://link.aps.org/doi/10.1103/42y2-bsh1}
}

\clearpage
\appendix
\markboth{Appendix}{Appendix}


\section{Replica computation for continuous embeddings}
\label{app:continuous_replica}

In this Appendix we derive the replica-symmetric equations used in
Sec.~\ref{sec:continuous_embeddings} to obtain the teacher-student overlap. We first recall the definitions and
normalization conventions of the continuous-embedding problem.

Each input is a sequence
\begin{equation}
    \bm\xi^\mu
    =
    \{\bm\xi_i^\mu\}_{i=1}^{N},
    \qquad
    \bm\xi_i^\mu\in\mathbb R^d,
\end{equation}
with the vectors $\bm\xi_i^\mu$ independently sampled from the sphere of
radius $1$. In particular,
\begin{equation}
    \left\langle
        \xi_{i,\gamma}^{\mu}
        \xi_{j,\delta}^{\nu}
    \right\rangle
    =
    \delta_{\mu\nu}
    \delta_{ij}
    \frac{\delta_{\gamma\delta}}{d}.
    \label{eq:app_input_covariance}
\end{equation}
We study the teacher-student setting, where the teacher weights $\mathbf W^\star \in \mathbb R^{N\times d\times d}$ are sampled uniformly on the sphere $\|\mathbf W^\star\|_F^2=Nd^2$ and the student weights $\mathbf W \in \mathbb R^{N\times d\times d}$ are constrained on the same sphere $\|\mathbf W\|_F^2=Nd^2$.

For a given input $\bm\xi^\mu$, the teacher and student pre-activation fields are
\begin{equation}
    \bm v^\mu
    =
    \frac{1}{\sqrt{Nd}}
    \sum_{i=1}^{N}
    \mathbf W_i^\star\bm\xi_i^\mu,
    \qquad
    \bm u^\mu
    =
    \frac{1}{\sqrt{Nd}}
    \sum_{i=1}^{N}
    \mathbf W_i\bm\xi_i^\mu,
\end{equation}
and the corresponding normalized outputs are
\begin{equation}
    \bm y^\mu
    =
    \frac{\bm v^\mu}{\|\bm v^\mu\|},
    \qquad \hat{\bm y}^\mu=\frac{\bm u^\mu}{\|\bm u^\mu\|}.
\end{equation}

Given a training set of $P=\alpha N$ patterns, the empirical loss has the
general form
\begin{equation}
    \mathcal L( W)
    =
    \sum_{\mu=1}^{P}
    \ell(\bm u^\mu, \bm y^\mu).
\end{equation}
We will focus on a specific sublcass of losses that only depend on the scalar product between the student pre-activation and the teacher output 
$h^\mu=\bm u^\mu\cdot\bm y^\mu$ and the student pre-activation squared norm  $\rho^\mu = \|\bm u^\mu\|^{2}$, 
\begin{equation}
    \mathcal L(\mathbf W)
    =
    \sum_{\mu=1}^{P}
    V(h^\mu,\rho^\mu).
\end{equation}
In particular, for the cross-entropy loss considered in the main text, the conditional distribution of a unit output
vector $\bm y$ given the student field $\bm u$ is
\begin{equation}
    p_\lambda(\bm y|\bm u)
    =
    \frac{
        \exp\left(
            \lambda\bm y\cdot\bm u
        \right)
    }{
        K_d(\lambda\|\bm u\|)
    },
    \label{eq:app_spherical_likelihood}
\end{equation}
where
\begin{equation}
    K_d(\lambda\|\bm u\|)
    =
    \int_{\mathbb S^{d-1}}
    d\omega(\bm y')
    \,
    \exp\left(
        \lambda\bm y'\cdot\bm u
    \right)
    \label{eq:app_spherical_partition}
\end{equation}
is the spherical partition function and $d\omega$ denotes the rotationally
invariant measure on the unit sphere. By rotational invariance, this integral
depends on $\bm u$ only through its norm. Choosing the polar axis along
$\bm u$, and defining
\begin{equation}
    c
    =
    \bm y'\cdot
    \frac{\bm u}{\|\bm u\|},
\end{equation}
one obtains
\begin{equation}
    K_d(\eta)
    =
    C_d
    \int_{-1}^{1}
    dc\,
    (1-c^2)^{\frac{d-3}{2}}
    e^{\eta c},
    \qquad
    \eta=\lambda\|\bm u\|,
\end{equation}
where $C_d$ is independent of $\eta$. Using the integral representation of the
modified Bessel function of the first kind $I_\nu$,
\begin{equation}
    K_d(\eta)
    \propto
    \eta^{1-d/2}
    I_{d/2-1}(\eta).
    \label{eq:app_spherical_partition_bessel}
\end{equation}
The proportionality constant is independent of the field and therefore
contributes only an additive constant to the loss.

The negative log-likelihood associated with
Eq.~\eqref{eq:app_spherical_likelihood} is
\begin{equation}
    -\log p_\lambda(\bm y|\bm u)
    =
    -\lambda\bm y\cdot\bm u
    +
    \log K_d(\lambda\|\bm u\|).
\end{equation}
Dividing by the positive constant $\lambda$, which does not change its
minimizer for fixed $\lambda>0$, and using
$h=\bm u\cdot\bm y$ and $\rho=\|\bm u\|^2$, we obtain
\begin{equation}
    V_{\rm CE}(h,\rho)
    =
    -h
    +
    \frac{1}{\lambda}
    \log K_d\!\left(
        \lambda\sqrt{\rho}
    \right).
    \label{eq:app_ce_loss}
\end{equation}
\subsection{Replicated partition function and order parameters}
The partition function at inverse temperature $\beta$ is defined as
\begin{equation}
    Z_\beta
    =
    \int d\varrho(\mathbf W)
    \exp\left[
        -\beta
        \sum_{\mu=1}^{P}
        V(h^\mu,\rho^\mu)
    \right],
\end{equation}
where $d\varrho(\mathbf W)$ is the uniform measure on the sphere
$\|\mathbf W\|_F^2=Nd^2$.
The corresponding averaged free energy is then defined as 
\begin{equation}
    f_\beta = -\frac{1}{\beta N} \,\overline{\log Z_\beta}\, ,
\end{equation}
where the overline denotes the average over the realization of the inputs $\bm \xi^\mu$ and the teacher. 
In order to compute the average of the logarithm we use the replica identity
$\overline{\log Z_\beta}=\lim_{n\to0}(\overline{Z_\beta^n}-1)/n$, and consider,
for integer $n$, the replicated partition function 
\begin{equation}
    \overline{Z_\beta^n}
    =
    \int
    \prod_{a=1}^{n}d\varrho(\mathbf W^a)
    \,
    \overline{
    \exp\left[
        -\beta
        \sum_{\mu=1}^{P}
        \sum_{a=1}^{n}
        V(h^{a,\mu},\rho^{a,\mu})
    \right]
    }.
    \label{eq:app_replicated_Z}
\end{equation}
where $\mathbf  W^a$, $a=1,\dots,n$, are $n$ replicas of the student weights and $h^{a,\mu}$ and $\rho^{a,\mu}$ are the quantities defined above evaluated for replica $a$

We will then obtain the averaged free energy as
\begin{equation}
    f_\beta =-\frac{1}{\beta N} \,\lim_{n\to0}\partial_n\log\overline{Z_\beta^n}\, .
\end{equation}
For a fixed value of the student and teacher weights, the pre-activation fields are sums of $N$ independent contributions and
become jointly Gaussian as $N\to\infty$ at fixed $d$. Their covariances are
\begin{align}
    \left\langle
        u_{\alpha}^{a,\mu}
        u_{\beta}^{b,\nu}
    \right\rangle
    &=
    \delta_{\mu\nu}
    \frac{Q_{\alpha\beta}^{ab}}{d},
    \\
    Q_{\alpha\beta}^{ab}
    &=
    \frac{1}{Nd}
    \sum_{i=1}^{N}
    \sum_{\gamma=1}^{d}
    W^{a}_{i,\alpha\gamma}
    W^{b}_{i,\beta\gamma},
    \label{eq:app_Q_def}
    \\
    \left\langle
        u_{\alpha}^{a,\mu}
        v_{\beta}^{\nu}
    \right\rangle
    &=
    \delta_{\mu\nu}
    \frac{R_{\alpha\beta}^{a}}{d},
    \\
    R_{\alpha\beta}^{a}
    &=
    \frac{1}{Nd}
    \sum_{i=1}^{N}
    \sum_{\gamma=1}^{d}
    W^{a}_{i,\alpha\gamma}
    W^{\star}_{i,\beta\gamma}.
    \label{eq:app_R_def}
\end{align}
The spherical constraint fixes
{\begin{equation}
    \operatorname{Tr}\mathbf Q^{aa}=d.
\end{equation}}
Because the input distribution and the teacher ensemble are rotationally
invariant in the embedding space, we look for a replica-symmetric (RS)
isotropic solution and make the ansatz
\begin{equation}
    Q_{\alpha\beta}^{ab}
    =
    \begin{cases}
        \delta_{\alpha\beta},
        & a=b,
        \\
        q\,\delta_{\alpha\beta},
        & a\neq b,
    \end{cases}
    \qquad
    R_{\alpha\beta}^{a}
    =
    r\,\delta_{\alpha\beta}.
    \label{eq:app_RS_ansatz}
\end{equation}

Introducing the teacher as replica $a=0$, {the block matrix of weight overlaps factorizes as}
{\begin{equation}
    \bm Q
    =
    \widetilde{\mathbf Q}\otimes\mathbb I_d,
    \label{eq:overlapMatrix}
\end{equation}}
with
\begin{equation}
    \widetilde{\bm Q}
    =
    \begin{pmatrix}
        1 & r & r & \cdots & r
        \\
        r & 1 & q & \cdots & q
        \\
        r & q & 1 & \cdots & q
        \\
        \vdots & \vdots & \vdots & \ddots & \vdots
        \\
        r & q & q & \cdots & 1
    \end{pmatrix}.
    \label{eq:app_Qtilde}
\end{equation}
and, consequently, 
\begin{equation}
    \langle\|\bm u^a\|^2\rangle=1,
    \qquad
    \langle\bm u^a\cdot\bm u^b\rangle=q,
    \qquad
    \langle\bm u^a\cdot\bm v\rangle=r.
\end{equation}

We now insert the definition of the overlap matrix in Eq.~\eqref{eq:overlapMatrix} inside the replicated partition function using the Dirac delta:
\begin{equation}
    1=\int\prod_{a\le b}dQ^{ab}\,
    \prod_{a\le b}\prod_{\alpha,\beta}
    \delta\!\left(
        Nd\,Q^{ab}_{\alpha\beta}
        -\sum_{i,\gamma}W^{a}_{i,\alpha\gamma}W^{b}_{i,\beta\gamma}
    \right),
    \qquad a,b=0,\dots,n .
\end{equation}
By reorganising the order of the terms, the replicated partition function splits into a term that counts the
weights compatible with given overlaps (\emph{the entropic term}) and a term that contains the contribution of the loss (\emph{the energetic term}) .
Since the $P$ patterns are independent, the latter appears as a $P$-th power:
\begin{equation}
    \overline{Z_\beta^n}
    =
    \int d\bm Q\;
    e^{N\mathcal S_n(\bm Q)}
    \left[\mathcal A_n(\bm Q)\right]^{P},
\end{equation}
with
\begin{align}
    \mathcal S_n(\bm Q)
    &=
    \frac1N\log\int\prod_{a=0}^{n}d\varrho(\bm W^a)\,
    \prod\delta(\cdots),
    \\
    \mathcal A_n(\bm Q)
    &=
    \int\prod_{a=0}^{n}d^d\bm u^a\;
    \mathcal N\!\left(\bm u^0,\dots,\bm u^n;\bm 0,{\bm Q/d}\right)
    \prod_{a=1}^{n}e^{-\beta V(h^a,\rho^a)},
\end{align}
where $\mathcal{N}(\{\bm u\}; \bm 0, \bm Q / d)$ denotes the Gaussian distribution of zero average and covariance $\bm Q / d$.
Within the RS ansatz in Eq.~\eqref{eq:app_RS_ansatz}, $\mathcal S_n=nG_S+\mathcal O(n^2)$ and
$\log\mathcal A_n=nG_E+\mathcal O(n^2)$. Remembering that $P=\alpha N$, this gives
\begin{equation}
    \overline{Z_\beta^n}
    \propto 
    \int dq\,dr\,
    \exp\left\{
        Nn
        \left[
            G_S(q,r)
            +
            \alpha G_E(q,r)
        \right]
    \right\},
    \label{eq:app_GS_GE}
\end{equation}
where $G_S$ is the entropic contribution and $G_E$ energetic one.

\subsection{Entropic contribution}

The term $\mathcal S_n(q,r)$ is the log-volume of weight configurations with a given replica overlap $q$ and student-teacher overlap $r$. For each replica $a$ and output index $\alpha$,we  collect the weights
into a vector
$\bm w^a_\alpha=(W^a_{i,\alpha\gamma})_{i,\gamma}\in\mathbb R^{Nd}$.
By Eq.~\eqref{eq:app_Q_def}, the Gram matrix (i.e., the matrix collecting all
pairwise scalar products) of these $(n+1)d$ vectors is
$Nd\,(\widetilde{\bm Q}\otimes\mathbb I_d)$. The volume of $k$ vectors in
$\mathbb R^{M}$ with Gram matrix $\mathbf G$ scales as $(\det \mathbf G)^{(M-k-1)/2}$ \cite{muirhead2009aspects}. With
$M=Nd\to\infty$ and $k=(n+1)d$ finite, and using
$\det(\widetilde{\bm Q}\otimes\mathbb I_d)=(\det\widetilde{\bm Q})^{d}$,
we obtain, up to terms independent of $q$ and $r$,
\begin{equation}
    G_S(q,r)
    =
    \frac{d^2}{2}
    \lim_{n\to0}
    \frac{1}{n}
    \log\det\widetilde{\bm Q}.
    \label{eq:app_GS_det}
\end{equation}
Equivalently, one can represent the $\delta$-functions through conjugate
matrices $\widehat{\bm Q}$, perform the Gaussian integrals over the $Nd^2$
weight components, and extremize over $\widehat{\bm Q}$; this leads to the
same result.

The determinant of Eq.~\eqref{eq:app_Qtilde} is
\begin{equation}
    \det\widetilde{\bm Q}
    =
    (1-q)^{n-1}
    \left[
        1+(n-1)q-nr^2
    \right].
\end{equation}
Expanding around $n=0$,
\begin{equation}
    \log\det\widetilde{\bm Q}
    =
    n
    \left[
        \log(1-q)
        +
        \frac{q-r^2}{1-q}
    \right]
    +
    \mathcal O(n^2),
\end{equation}
we get 
\begin{equation}
    G_S(q,r)
    =
    \frac{d^2}{2}
    \left[
        \log(1-q)
        +
        \frac{q-r^2}{1-q}
    \right].
    \label{eq:app_GS_final}
\end{equation}

\subsection{Energetic contribution}

In order to compute the energetic contribution, we rewrite the preactivations $\bm u^a$ in terms of  independent cavity variables. We introduce
\begin{equation}
    \bm v_0,\bm z,\bm x^a
    \sim
    \mathcal N(\bm 0,\mathbb I_d)
\end{equation}
where $\bm v=\bm v_0/\sqrt d$ is the teacher pre-activation, $\bm z$ is a stochasticity shared by all the replicas, and $\bm x^a$ is a private noise and is independent from one replica to another. Then by writing
\begin{equation}
        \bm u^a
        =
        \frac{1}{\sqrt d}\left(
        r\bm v_0
        +
        \sqrt{q-r^2}\,\bm z
        +
        \sqrt{1-q}\,\bm x^a
        \right)
    \label{eq:app_cavity_decomposition}
\end{equation}
one correctly reproduces all the covariances $Q^{ab}/d$ and $R^a/d$ associated with Eq.~\eqref{eq:app_RS_ansatz}.

Denoting by $\mathcal D\bm x$
the Gaussian measure $\mathcal N(0,\mathbf I_d)$ on $\mathbb R^d$, conditioning on
$(\bm v_0,\bm z)$ makes the $n$ student replicas independent, and therefore
\begin{align}
    \mathcal A_n
    &=
    \mathbb E_{{\bm v_0},\bm z}
    \left[
        \left(
            \int
            \mathcal D\bm x\,
            e^{-\beta V(h,\rho)}
        \right)^n
    \right]
    \nonumber\\
    &=
    1
    +
    n\,
    \mathbb E_{{\bm v_0},\bm z}
    \left[
        \log
        \int
        \mathcal D\bm x\,
        e^{-\beta V(h,\rho)}
    \right]
    +
    \mathcal O(n^2).
\end{align}
Hence
\begin{equation}
    G_E(q,r)
    =
    \int
    \mathcal D{\bm v_0}\,
    \mathcal D\bm z\,
    \log
    \left[
        \int
        \mathcal D\bm x\,
        e^{-\beta V(h,\rho)}
    \right].
    \label{eq:app_GE_vector}
\end{equation}

For a given teacher field, we use rotational invariance to rotate the
coordinate system so that the normalized teacher direction is aligned with
the first canonical basis vector,
\begin{equation}
    \bm y
    =
    \frac{\bm v}{\|\bm v\|}
    =
    \bm e_1.
\end{equation}
In this basis we decompose every vector into its component along $\bm y$
and its component in the orthogonal $(d-1)$-dimensional subspace,
\begin{equation}
    \bm u
    =
    (u_1,\bm u_\perp),
    \qquad
    \bm z
    =
    (z_1,\bm z_\perp),
    \qquad
    \bm u_\perp,\bm z_\perp
    \in
    \mathbb R^{d-1},
\end{equation}
while
\begin{equation}
    \bm v_0=(v,\bm 0),
    \qquad
    \bm v=(v/\sqrt d,\bm 0)
\end{equation}
by construction. Writing
\begin{equation}
    v=\|\bm v_0\|=\sqrt d\,\|\bm v\|,
    \qquad
    h=u_1=\bm u\cdot\bm y,
    \qquad
    s=\|\bm u_\perp\|^2,
\end{equation}
one has, denoting by $\chi_k$ and $\chi_k^2$ the chi and chi-square
distributions with $k$ degrees of freedom,
\begin{equation}
    v\sim\chi_d,
    \qquad
    p
    \equiv
    \|\bm z_\perp\|^2
    \sim
    \text{CHI}^2_{d-1},
    \qquad
    z
    \equiv
    z_1
    \sim
    \mathcal N(0,1).
\end{equation}

The longitudinal component of
Eq.~\eqref{eq:app_cavity_decomposition} is
\begin{equation}
    h
    =
    \frac{rv+\sqrt{q-r^2}\,z+\sqrt{1-q}\,x_1}{\sqrt d},
\end{equation}
and therefore, conditioned on the values of $v$ and $z$,
{\begin{equation}
    h
    \sim
    \mathcal N
    \left(
        (rv+\sqrt{q-r^2}\,z)/\sqrt d,\,
        (1-q)/d
    \right).
    \label{eq:app_h_distribution}
\end{equation}}
For the transverse component, conditioned on $\bm z_\perp$,
\begin{equation}
    \bm u_\perp
    =
    \frac{1}{\sqrt d}\left(
    \sqrt{q-r^2}\,\bm z_\perp
    +
    \sqrt{1-q}\,\bm x_\perp\right).
\end{equation}
This is therefore a $(d-1)$-dimensional Gaussian
vector with covariance $(1-q)\mathbb I_{d-1}/d$ and squared mean norm
\begin{equation}
    L
    =
    \frac{q-r^2}{d}
    \|\bm z_\perp\|^2
    =
    (q-r^2)p/d.
\end{equation}
Consequently, remembering that $s = \|\bm u_\perp\|^2$, we have
\begin{equation}
    \frac{s}{\Delta}
    \sim
    \text{CHI}_{d-1}^{\prime\,2}
    \left(
        \frac{L}{\Delta}
    \right),
    \qquad
    \Delta=(1-q)/d,
    \label{eq:app_s_distribution}
\end{equation}
where $\text{CHI}_k^{\prime\,2}(\Lambda)$ denotes a non-central chi-square
distribution with $k$ degrees of freedom and dimensionless noncentrality
parameter $\Lambda$.

We denote directly by
$f_{\chi^2}(s;k,L,\Delta)$ the probability density of the physical variable
$s$ 
\begin{equation}
    f_{\chi^2}(s;k,L,\Delta)
    =
    \frac{1}{2\Delta}
    \left(
        \frac{s}{L}
    \right)^{\frac{k-2}{4}}
    \exp
    \left[
        -\frac{s+L}{2\Delta}
    \right]
    I_{\frac{k}{2}-1}
    \left(
        \frac{\sqrt{sL}}{\Delta}
    \right).
    \label{eq:app_noncentral_chi}
\end{equation}
For vanishing noncentrality $L=0$, this reduces to
\begin{equation}
    f_{\chi^2}(s;k,0,\Delta)
    =
    \frac{
        s^{k/2-1}
        e^{-s/(2\Delta)}
    }{
        (2\Delta)^{k/2}
        \Gamma(k/2)
    }.
\end{equation}

The energetic term can then be written as
{\begin{align}
    G_E(q,r)
    &=
    \Bigg\langle
        \log
        \int dh\,ds\,
        \mathcal N
        \left(
            h;
            (rv+\sqrt{q-r^2}\,z)/\sqrt d,\,
            (1-q)/d
        \right)
        \nonumber\\
    &\hspace{2.0cm}\times
        f_{\chi^2}
        \left(
            s;
            d-1,
            (q-r^2)p/d,
            (1-q)/d
        \right)
        e^{-\beta V(h,h^2+s)}
    \Bigg\rangle_{v,z,p}.
    \label{eq:app_GE_scalar}
\end{align}}
Here the bracket average $\langle \cdot  \rangle $ denotes
\begin{align}
    \langle {A}\rangle_{v,z,p}
    &=
    \int_0^\infty dv\,
    \frac{
        2^{1-d/2}
    }{
        \Gamma(d/2)
    }
    v^{d-1}e^{-v^2/2}
    \nonumber\\
    &\qquad\times
    \int Dz
    \int_0^\infty dp\,
    f_{\chi^2}(p;d-1,0,1)\,
    {A(v,z,p)},
    \label{eq:app_average_definition}
\end{align}
with $Dz =\frac{dz}{\sqrt{2\pi}}\, e^{-z^2/2}. $

\subsection{Zero-temperature limit}

Since we are interested in the minima of the empirical loss, we will consider the limit $\beta\to\infty$, where only the minimum energy student configurations contribute to the free energy. We assume that the potential $V(h, \rho)$ has a unique minimizer, so that at zero temperature two equilibrium replicas approach the same point, and the following scaling for the replica overlap
\begin{equation}
    q\to1,
    \qquad
    \chi
    =
    \beta(1-q)
    =
    \mathcal O(1).
    \label{eq:app_zeroT_scaling}
\end{equation}

The entropic term then converges to
\begin{equation}
    \frac{1}{\beta}G_S(q,r)
    \longrightarrow
    \frac{d^2}{2\chi}(1-r^2).
    \label{eq:app_GS_zeroT}
\end{equation}

The longitudinal Gaussian in Eq.~\eqref{eq:app_h_distribution} has the
large-$\beta$ behavior
\begin{equation}
    \mathcal N
    \left(
        h;
        (rv+\sqrt{1-r^2}\,z)/\sqrt d,
        \frac{\chi}{d\beta}
    \right)
    \propto 
    \exp\left[
        -\frac{d\beta}{2\chi}
        \left(
            h-(rv+\sqrt{1-r^2}\,z)/\sqrt d
        \right)^2
    \right],
    \label{eq:app_h_pdf}
\end{equation}
while for the transverse term, using the large-argument expansion
$I_\nu(x)\simeq e^{x}/\sqrt{2\pi x}$ in Eq.~\eqref{eq:app_noncentral_chi}, gives
\begin{equation}
    f_{\chi^2}
    \left(
        s;
        d-1,
        p(1-r^2)/d,
        \frac{\chi}{d\beta}
    \right)
    \asymp
    \exp\left[
        -\frac{d\beta}{2\chi}
        \left(
            \sqrt{s}
            -
            \sqrt{p(1-r^2)/d}
        \right)^2
    \right].
    \label{eq:app_chi_zeroT}
\end{equation}
It is convenient to introduce
\begin{equation}
    t=\sqrt{s},
    \qquad
    \bar h
    =
    (rv+\sqrt{1-r^2}\,z)/\sqrt d,
    \qquad
    \bar t
    =
    \sqrt{1-r^2}\sqrt{p/d}.
\label{eq:app_relations_t_hbar_tbar}
\end{equation}
The remaining integral of the energetic term can then be evaluated by Laplace's method. For fixed
$(v,z,p)$ we define
\begin{equation}
    \mathcal M
    \left(
        h,t;\chi,r,v,z,p
    \right)
    =
    V(h,h^2+t^2)
    +
    \frac{d}{2\chi}
    \left[
        (h-\bar h)^2
        +
        (t-\bar t)^2
    \right],
    \qquad
    t\geq0,
    \label{eq:app_M_definition}
\end{equation}
and
\begin{equation}
    (h_\star,t_\star)
    =
    \operatorname*{arg\,min}_{h\in\mathbb R,\;t\geq0}
    \mathcal M(h,t;\chi,r,v,z,p).
    \label{eq:app_inner_minimum}
\end{equation}
Whenever this proximal problem has a unique minimizer, as in the
finite-concentration spherical cross-entropy case considered in the main
text, Laplace's method gives
\begin{equation}
    \frac{1}{\beta}G_E
    \longrightarrow
    -
    \left\langle
        \mathcal M(h_\star,t_\star)
    \right\rangle_{v,z,p}.
\end{equation}
The zero-temperature free energy $\Phi_0=\lim_{\beta\to\infty}f_\beta$ is therefore
\begin{equation}
    \Phi_0(\chi,r)
    =
    -
    \frac{d^2}{2\chi}(1-r^2)
    +
    \alpha
    \left\langle
        \mathcal M(h_\star,t_\star)
    \right\rangle_{v,z,p},
    \label{eq:app_zeroT_functional}
\end{equation}
whose value at the saddle point is the minimal training loss per input
dimension.

\subsection{Saddle-point equations}

In order to obtain the values $(\chi,r)$ corresponding to the training loss minimizers, we study the stationary points of
$\Phi_0(\chi,r)$ in Eq.~\eqref{eq:app_zeroT_functional}. Since
$(h_\star,t_\star)$ minimizes $\mathcal M$, the derivatives of $\Phi_0$ act only on
the explicit dependence on $\chi$ and $r$, and not on $(h_\star,t_\star)$. The
derivative with respect to $\chi$ is
\begin{equation}
    \frac{\partial\Phi_0}{\partial\chi}
    =
    \frac{d^2}{2\chi^2}(1-r^2)
    -
    \frac{\alpha d}{2\chi^2}
    \left\langle
        (h_\star-\bar h)^2
        +
        (t_\star-\bar t)^2
    \right\rangle_{v,z,p},
\end{equation}
and setting it to zero gives
\begin{equation}
    d(1-r^2)
    =
    \alpha
    \left\langle
        (h_\star-\bar h)^2
        +
        (t_\star-\bar t)^2
    \right\rangle_{v,z,p}.
    \label{eq:app_saddle_chi}
\end{equation}
The derivative with respect to $r$ is
\begin{equation}
    \frac{\partial\Phi_0}{\partial r}
    =
    \frac{d^2 r}{\chi}
    -
    \frac{\alpha d}{\chi}
    \left\langle
        (h_\star-\bar h)\,\frac{\partial\bar h}{\partial r}
        +
        (t_\star-\bar t)\,\frac{\partial\bar t}{\partial r}
    \right\rangle_{v,z,p},
\end{equation}
with
\begin{equation}
    \frac{\partial\bar h}{\partial r}
    =
    \frac{1}{\sqrt d}\left(v-\frac{rz}{\sqrt{1-r^2}}\right),
    \qquad
    \frac{\partial\bar t}{\partial r}
    =
    -\frac{r}{\sqrt{1-r^2}}\sqrt{\frac{p}{d}}.
\end{equation}
Setting it to zero and multiplying by $\sqrt{d(1-r^2)}$ gives
\begin{equation}
    r\, d^{3/2}\sqrt{1-r^2}
    =
    \alpha
    \left\langle
        (h_\star-\bar h)
        \left(
            v\sqrt{1-r^2}-rz
        \right)
        +
        (\bar t-t_\star)\,r\sqrt p
    \right\rangle_{v,z,p}.
    \label{eq:app_saddle_r}
\end{equation}
Equations~\eqref{eq:app_saddle_chi} and \eqref{eq:app_saddle_r}, together
with the inner minimization Eq.~\eqref{eq:app_inner_minimum}, can be solved numerically: for each trial pair $(r,\chi)$ and each quadrature point
$(v,z,p)$ we compute $(h_\star,t_\star)$, insert them in the two averages,
and solve the resulting equations jointly for $(r,\chi)$.

We note that for the spherical cross-entropy loss, Eq.~\eqref{eq:app_ce_loss}, the
stationarity condition of $\mathcal M$ reads
\begin{equation}
    \left[1+\frac{\chi}{d}\,\frac{B_d(\lambda R_\star)}{R_\star}\right]
    (h_\star,t_\star)
    =
    \left(\bar h+\frac{\chi}{d},\,\bar t\right),
    \qquad
    B_d(x)=\frac{I_{d/2}(x)}{I_{d/2-1}(x)},
\end{equation}
with $R_\star=\sqrt{h_\star^2+t_\star^2}$. Therefore $(h_\star,t_\star)$ is parallel
to $(\bar h+\chi/d,\,\bar t)$, and its norm is the unique non-negative solution of
\begin{equation}
    R_\star+\frac{\chi}{d}\,B_d(\lambda R_\star)
    =
    \sqrt{\left(\bar h+\frac{\chi}{d}\right)^2+\bar t^2}.
    \label{eq:app_continuous_ce_prox}
\end{equation}

\subsection{Large embedding dimension}
\label{sec:continuous_large_d}

We now want to study the zero-temperature saddle-point equations in the
regime of large embedding dimension $d\to\infty$. Since this limit is
taken after the saddle-point evaluation, we still have $d\ll N$.
For $d\gg 1$, the two variables $v\sim\chi_d$ and $\sqrt p\sim\chi_{d-1}$
concentrate around $\sqrt d$, while $z\sim\mathcal N(0,1)$ remains of order
one. Consequently, the cavity fields become deterministic,
$\bar h\to r$ and $\bar t\to\sqrt{1-r^2}$, and the saddle-point equations,
Eqs.~\eqref{eq:app_saddle_chi} and \eqref{eq:app_saddle_r}, divided by $d$,
converge to
\begin{align}
    1-r^2
    &=
    \widetilde\alpha
    \left[
        (h_\star-r)^2+\bigl(t_\star-\sqrt{1-r^2}\bigr)^2
    \right],
    \label{eq:large_d_chi_raw}
    \\
    r\sqrt{1-r^2}
    &=
    \widetilde\alpha
    \left[
        \sqrt{1-r^2}\,(h_\star-r)
        +
        r\bigl(\sqrt{1-r^2}-t_\star\bigr)
    \right],
    \label{eq:large_d_r_raw}
\end{align}
where we introduced the rescaled load $\widetilde\alpha=\alpha/d$.

For the cross-entropy loss, $(h_\star,t_\star)$ is given by
Eq.~\eqref{eq:app_continuous_ce_prox}, which in the same limit reads
\begin{equation}
    (h_\star,t_\star)
    =
    \frac{R_\star}{\sqrt{1+2r\widetilde\chi+\widetilde\chi^2}}
    \left(r+\widetilde\chi,\,\sqrt{1-r^2}\right),
    \qquad
    R_\star+\widetilde\chi\,B_d(\lambda R_\star)
    =
    \sqrt{1+2r\widetilde\chi+\widetilde\chi^2},
    \label{eq:large_d_prox}
\end{equation}
where $\chi$ enters only through $\widetilde\chi=\chi/d$. From the integral
representation of $K_d$,
\begin{equation}
    B_d(x)
    =\frac{\mathrm d}{\mathrm dx}\log K_d(x)
    =
    \frac{\int_{-1}^{1}du\,u\,(1-u^2)^{\frac{d-3}{2}}e^{xu}}
    {\int_{-1}^{1}du\,(1-u^2)^{\frac{d-3}{2}}e^{xu}}.
\end{equation}
For fixed $x$, the weight is concentrated around $u=0$ with variance $1/d$,
so that $B_d(x)\simeq x/d\to0$. For $x=yd$, with $y=\mathcal O(1)$, the
integrals are instead dominated by the saddle point of
$yu+\tfrac12\log(1-u^2)$, i.e.\ $y=u/(1-u^2)$, which gives
\begin{equation}
    B_d(dy)
    \;\longrightarrow\;
    \widetilde B(y)
    =
    \frac{2y}{1+\sqrt{1+4y^2}}.
    \label{eq:large_d_bessel}
\end{equation}
The loss normalization therefore contributes at large $d$ through
$\widetilde\lambda=\lambda/d$, and in Eq.~\eqref{eq:large_d_prox} we can
replace $B_d(\lambda R_\star)\to\widetilde B(\widetilde\lambda R_\star)$.
Inserting Eq.~\eqref{eq:large_d_prox} into Eqs.~\eqref{eq:large_d_chi_raw}
and \eqref{eq:large_d_r_raw}, the bracket in the second equation simplifies
to $\sqrt{1-r^2}\,\widetilde\chi\,R_\star/\sqrt{1+2r\widetilde\chi+\widetilde\chi^2}$,
and we obtain
\begin{equation}
    \begin{aligned}
    1-r^2
    &=
    \widetilde\alpha
    \left[
        \left(
            \frac{R_\star\,(r+\widetilde\chi)}{\sqrt{1+2r\widetilde\chi+\widetilde\chi^2}}-r
        \right)^2
        +
        \left(
            1-\frac{R_\star}{\sqrt{1+2r\widetilde\chi+\widetilde\chi^2}}
        \right)^2(1-r^2)
    \right],
    \\
    r
    &=
    \widetilde\alpha\,\widetilde\chi\,
    \frac{R_\star}{\sqrt{1+2r\widetilde\chi+\widetilde\chi^2}},
    \end{aligned}
    \label{eq:large_d_ce_saddle}
\end{equation}
with $R_\star$ the solution of
$R_\star+\widetilde\chi\,\widetilde B(\widetilde\lambda R_\star)
=\sqrt{1+2r\widetilde\chi+\widetilde\chi^2}$.
These equations depend on $d$ only through the rescaled variables
\begin{equation}
    \widetilde\alpha=\frac{\alpha}{d},
    \qquad
    \widetilde\chi=\frac{\chi}{d},
    \qquad
    \widetilde\lambda=\frac{\lambda}{d},
    \label{eq:large_d_scales}
\end{equation}
and admit solutions with $0<r<1$ and finite $\widetilde\chi$ for finite
$\widetilde\alpha$, which confirms a posteriori the scalings
$\alpha,\chi,\lambda\propto d$. 


\subsection{Generalization error}

We now connect the teacher-student overlap $r$ to the generalisation performances of the student network on unseen inputs.
For a new pattern, the student and teacher pre-activation fields $(\bm u, \bm v)$ are jointly Gaussian with
\begin{equation}
    \langle v_\alpha v_\beta\rangle
    =
    \langle u_\alpha u_\beta\rangle
    =
    \delta_{\alpha\beta}/d,
    \qquad
    \langle u_\alpha v_\beta\rangle
    =
    r\,\delta_{\alpha\beta}/d.
\end{equation}
We define their normalized overlap as
\begin{equation}
    \widehat{\Delta}
    =
    \frac{
        \bm u\cdot\bm v
    }{
        \|\bm u\|\,\|\bm v\|
    } = \bm y \cdot \hat{\bm y}
\end{equation}
and the corresponding generalisation error as
\begin{equation}
    \epsilon_g(r)
    =
    \frac12
    \left[
        1-
        \langle\widehat{\Delta}\rangle
    \right],
\end{equation}
where $\langle \cdot \rangle $ denotes the average over the input distribution.

The first moment of the angle between two correlated isotropic Gaussian vectors
is
\begin{equation}
    \left\langle\widehat{\Delta}\right\rangle
    =
    r\,
    \frac{A_d^2}{d}
    \,
    {}_2F_1
    \left(
        \frac12,\frac12;
        \frac d2+1;
        r^2
    \right),
\end{equation}
where
$    A_d
    =
    \sqrt{2}
        \Gamma\!\left(\frac{d+1}{2}\right)
        /
        \Gamma\!\left(\frac d2\right)
    $
and ${}_2F_1$ is the Gauss hypergeometric function. We can therefore write the generalisation error as a function of the student-teacher overlap as
\begin{equation}
    \epsilon_g(r)
    =
    \frac12
    \left[
        1
        -
        r\frac{A_d^2}{d}
        {}_2F_1
        \left(
            \frac12,\frac12;
            \frac d2+1;
            r^2
        \right)
    \right].
    \label{eq:app_generalization_error}
\end{equation}

For $d=1$,
\begin{equation}
    A_1^2=\frac{2}{\pi},
    \qquad
    r\,{}_2F_1
    \left(
        \frac12,\frac12;
        \frac32;
        r^2
    \right)
    =
    \arcsin r,
\end{equation}
and we recover the classical scalar perceptron dependence
\begin{equation}
    \epsilon_g^{d=1}(r)
    =
    \frac{1}{\pi}\arccos r.
\end{equation}
On the other hand, for $d\to\infty$, one has
\begin{equation}
    \frac{A_d^2}{d}
    \longrightarrow1,
    \qquad
    {}_2F_1
    \left(
        \frac12,\frac12;
        \frac d2+1;
        r^2
    \right)
    \longrightarrow1
\end{equation} 
and therefore in the $d\gg1$ regime
\begin{equation}
    \epsilon_g(r)
    =
    \frac{1-r}{2}
    +
    \mathcal O(d^{-1}).
\end{equation}



\subsection{Memorization capacity}
\label{app:continuous_classification_capacity}

A complementary limit of the same theory is obtained by removing the
teacher--student correlation and considering random labels. In replica
language this corresponds to setting
\begin{equation}
    r=0.
\end{equation}
The resulting version-space volume measures the number of student weights that
can satisfy a set of unrelated input--output constraints and therefore gives a
memorization, rather than a generalization, capacity.

We consider hard constraints: denoting by $\Theta$ the Heaviside step
function, the potential is $V=\Theta(\cdot)$, so that at zero temperature
$e^{-\beta V}\to\mathbf 1_{\mathcal C}(h,t)$, the indicator of the set
$\mathcal C$ of admissible fields. $Z_\infty$ is then the Gardner volume of
weights satisfying all $P$ constraints. We work in the regime
$\alpha<\alpha_c$, where this volume is non-empty: as $\alpha$ increases it
shrinks and the overlap $q$ between two solutions grows, until at
$\alpha\to\alpha_c^-$ only a single solution survives and $q\to1$. The
capacity is therefore obtained from the saddle point of
Eq.~\eqref{eq:app_GS_GE} at $r=0$, in the limit $q\to1$.

In this limit the entropic term Eq.~\eqref{eq:app_GS_final} is dominated by
\begin{equation}
    G_S(q,0)\simeq\frac{d^2}{2(1-q)}.
\end{equation}
In the energetic term Eq.~\eqref{eq:app_GE_scalar}, {we use standardized coordinates $H=\sqrt d\,h$ and $T=\sqrt d\,t$. Their Gaussian fluctuations have variance $1-q$ around the standardized cavity point $(z,\bar t)$, where $z\sim\mathcal N(0,1)$ and $\bar t=\sqrt p\sim\chi_{d-1}$ are independent. In the following distance calculation only, $\mathcal C$ denotes the admissible set in {$(H,T)$} coordinates, and its projection threshold is $K=\sqrt d\,\kappa$. Using Eq.~\eqref{eq:app_chi_zeroT} with $\chi/\beta$ replaced by $1-q$ after this coordinate rescaling}, the integral over the admissible fields becomes, up to
sub-exponential prefactors,
{\begin{equation}
    \int_{\mathcal C} dH\,dT\,
    \exp\left[
        -\frac{(H-z)^2+(T-\bar t)^2}{2(1-q)}
    \right]
    \asymp
    \exp\left[
        -\frac{D_{\mathcal C}^2(z,\bar t)}{2(1-q)}
    \right],
\end{equation}}
with
{\begin{equation}
\label{eq:D_def}
    D_{\mathcal C}^2(z,\bar t)
    =
    \min_{(H,T)\in\mathcal C}
    \left[(H-z)^2+(T-\bar t)^2\right],
\end{equation}}
since as $q\to1$ the fluctuations vanish and the integral is dominated by the
admissible point closest to $(z,\bar t)$. $D_{\mathcal C}^2$ is thus the
squared distance from the cavity point to the admissible set, and it vanishes
when the cavity point is already admissible. Taking the logarithm and
averaging,
\begin{equation}
    G_E(q,0)\simeq-\frac{\left\langle D_{\mathcal C}^2(z,\bar t)\right\rangle_{z,\bar t}}{2(1-q)}.
\end{equation}
Both terms diverge as $(1-q)^{-1}$, and
\begin{equation}
    G_S+\alpha G_E
    \simeq
    \frac{d^2-\alpha\left\langle D_{\mathcal C}^2\right\rangle_{z,\bar t}}{2(1-q)}.
\end{equation}
The saddle point can reach $q\to1$ only if the coefficient of the divergence
vanishes, which fixes the critical load through
\begin{equation}
    d^2
    =
    \alpha_c
    \left\langle D_{\mathcal C}^2(z,\bar t)\right\rangle_{z,\bar t}.
    \label{eq:app_capacity_condition}
\end{equation}
The capacity is thus determined by the geometry of $\mathcal C$ in the
{$(H,T)$} plane. In what follows, we evaluate it for the cases of standard and vector alignments of Eqs~\eqref{eq:continuous_gardner_capacity},\eqref{eq:continous_alignment_capacity} (SA and VA, recall).
Consider first SA:
\begin{equation}
    V_\kappa(h)
    =
    \Theta(\kappa-h),
\end{equation}
which at zero temperature becomes $e^{-\beta V_\kappa}\to\Theta(h-\kappa)$:
every stored pattern must satisfy
\begin{equation}
    h\geq \kappa.
\end{equation}
\begin{figure}
    \centering
    \includegraphics[width=0.9\linewidth]{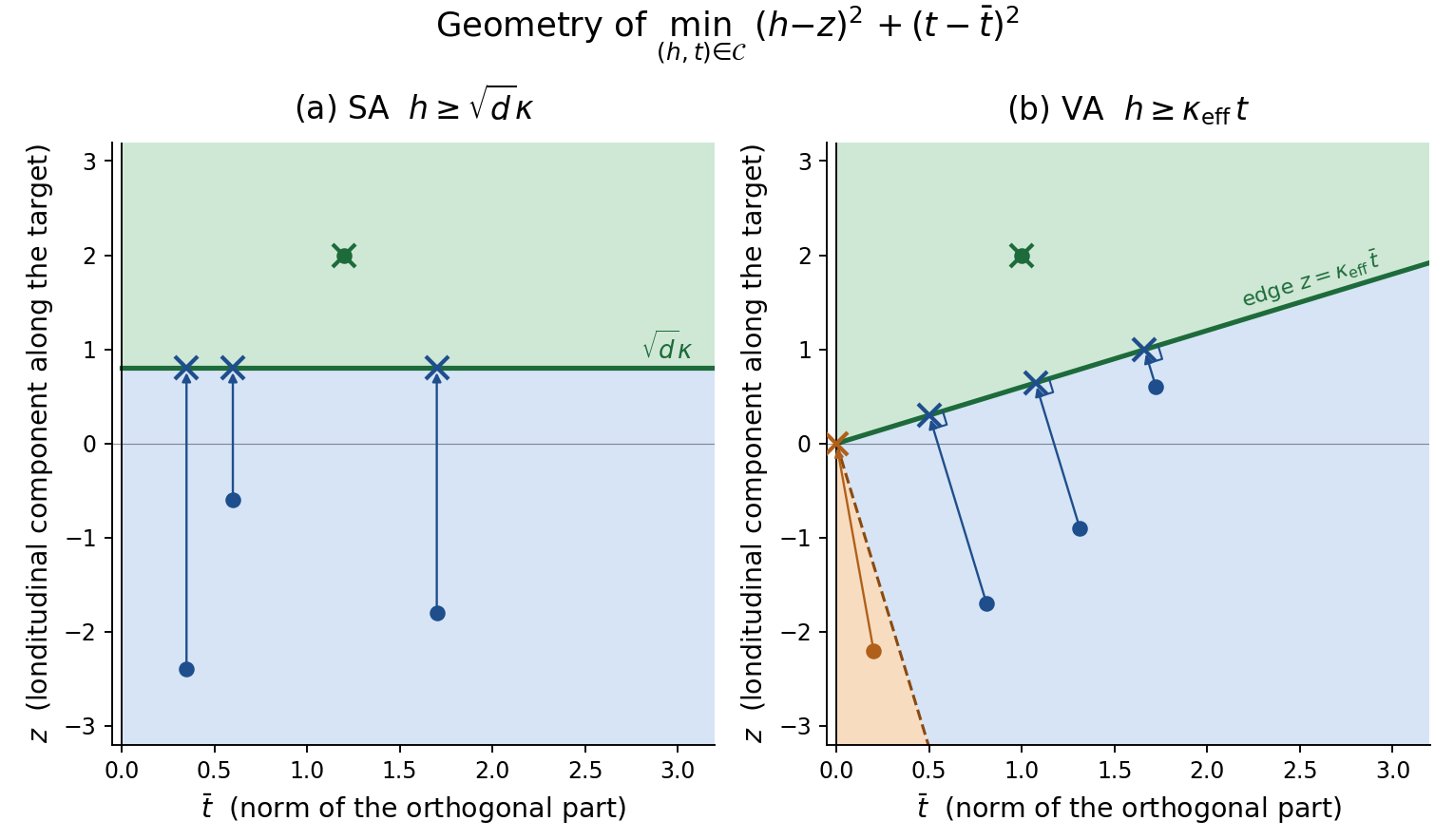}
    \caption{\textbf{Schematic representation of the minimization of the potential}. 
    We show how the square distance to points $(z, \bar{t})$ is minimised under the different constraints of SA and VA storage.
    \underline{Left panel}: the minimisation for SA storage. For different points (full dots) in the $z, \bar{t}$ half-plane, we show the projections (crosses) on the satisfability half-plane $z>\sqrt{d}\kappa$, indicated in green, corresponding to the points $h_*, t_*$ that minimise the potential on the constraint set $\mathcal{C}=\{h\in \mathbb{R}, t\geq 0: h\geq \sqrt{d}\kappa \}$. \underline{Right panel}: the minimisation for VA storage. Again, points in the $z, \bar{t}$ plane (full symbols) are associated to the corresponding points $h_*, t_*$ that minimise the potential on the constraint set $\mathcal{C}=\{h\in \mathbb{R}, t\geq 0: h\geq \kappa_{\rm eff}t\}$ (crosses). In the blue region $\mathcal{C}=\{h\in \mathbb{R}, t\geq 0: -\frac{t}{\kappa_{\rm eff}}\leq h\leq \kappa_{\rm eff}t\}$, to minimise the potential each point is projected onto the line $z=\kappa_{\rm eff} t$. In the orange region ($z<-\frac{\bar{t}}{\kappa_{\rm eff}}$), all points are projected to the origin.}
    \label{fig:scheme_minimization_V}
\end{figure}
The standardized admissible set $\mathcal C=\{(H,T):\,H\geq K\}$ constrains only $H$, so the closest admissible point is $(\max(z,K),\bar t)$, with $K=\sqrt d\,\kappa$ (see the left panel in Fig.~\ref{fig:scheme_minimization_V}). Hence $D_{\mathcal C}^2=(K-z)_+^2$, with $(x)_+=\max(x,0)$, and
Eq.~\eqref{eq:app_capacity_condition} gives
{\begin{equation}
    d^2
    =
    \alpha_c^{\rm (SA)}
    \int_{-\infty}^{\sqrt d\,\kappa}
    Dz\,
    (z-\sqrt d\,\kappa)^2,
\end{equation}}
that is
\begin{equation}
    \alpha_c^{\rm (SA)}(\kappa;d)
    =
    \frac{
        d^2
    }{
        \displaystyle
        \int_{-\infty}^{\sqrt d\,\kappa}
        Dz\,
        (z-\sqrt d\,\kappa)^2
    }
    \equiv
    d^2\alpha_c^{\rm (G)}(\sqrt d\,\kappa).
    \label{eq:app_gardner_longitudinal}
\end{equation}
In particular, at zero margin,
\begin{equation}
    \alpha_c^{\rm (SA)}(0;d)=2d^2.
\end{equation}

Finally, we discuss VA. Recalling that $h$ and $t$ are respectively the longitudinal and orthogonal components of the student preactivation, the potential reads
\begin{equation}
    V_\kappa(h, t)
    =
    \Theta\left(\kappa-\frac{h}{\sqrt{h^2+t^2}}\right).
\end{equation}
At zero temperature, every stored pattern must abide by the following inequality
\begin{equation}
    h \geq \kappa_{\rm eff}(\kappa)t\qquad \kappa_{\rm eff}(\kappa) = \frac{\kappa}{\sqrt{1-\kappa^2}}\qquad 0\leq \kappa<1.
\end{equation}
We can evaluate the minimization of the distance $\mathcal{D}_{\mathcal{C}}^2$ of Eq.~\eqref{eq:D_def} on the set {$\mathcal{C}=\{H\in \mathbb{R}, T\geq 0: H\geq \kappa_{\rm eff} T\}$} by considering three different regions of the $(z,\bar{t})$ half-plane, as shown in the right panel of Fig.~\ref{fig:scheme_minimization_V}.
\begin{itemize}
    \item Region $z\geq\kappa_{\rm eff} \bar{t}$: here, by definition, every constraint is satisfied. The point of minimum is trivially {$H_*=z$,$T_*=\bar{t}$} and the distance \eqref{eq:D_def} is identically zero.
    \item Region $-\frac{\bar{t}}{\kappa_{\rm eff}}\leq z<\kappa_{\rm eff} \bar{t}$: here, the point of minimum on $\mathcal{C}$ is obtained by projecting $(z, \bar{t})$ onto the line {$H=\kappa_{\rm eff} T$}; one can easily obtain {$H_*=z+\frac{\kappa_{\rm eff}\bar t -z}{1+\kappa_{\rm eff}^2}$}, {$T_*=\bar{t}-\frac{\kappa_{\rm eff}(\kappa_{\rm eff}\bar t -z)}{1+\kappa_{\rm eff}^2}$} and thus the distance reads $\mathcal{D}_\mathcal{C}^2(z, \bar{t}) = \frac{(\kappa_{\rm eff}\bar{t}-z)^2}{1+\kappa_{\rm eff}^2}$.
    \item Region $z<-\frac{\bar{t}}{\kappa_{\rm eff}}$: below this line, the point closest to $(z, \bar{t})$ in $\mathcal{C}$ is always the origin. Thus, {$H_*=T_*=0$} and $\mathcal{D}_\mathcal{C}^2(z, \bar{t}) = z^2 + \bar{t}^2$.
\end{itemize}

This returns the squared distance in Eq.~\eqref{eq:continous_alignment_capacity}, which we also report here for reader's convenience:
\begin{equation*}
\begin{gathered}
    \,\\
    \alpha_c^{\rm (VA)}(\kappa;d)
    =
    \frac{
        d^2
    }{
        \displaystyle
        \mathbb E_{
            z\sim\mathcal N(0,1),
            \,
            \bar t\sim\chi_{d-1}
        }
        \left[
            D_\kappa^2(z,\bar t)
        \right]
    } \\ \\ \text{with} \\ \\
    \;D_\kappa^2(z,\bar t)
    =
    \frac{(z-\kappa_{\rm eff}\bar{t})^2}{1+\kappa_{\rm eff}^2}\Theta(\kappa_{\rm eff}\bar{t}-z)\Theta\left(z+\frac{\bar{t}}{\kappa_{\rm eff}}\right)+(z^2+\bar{t}^2)\Theta\left(-z-\frac{\bar{t}}{\kappa_{\rm eff}}\right).\; \\
    \,
\end{gathered}
\end{equation*}
The integrals in the equations of the critical capacity for SA and VA admit closed-form expressions:
\begin{equation}
    \begin{split}
        \int_{-\infty}^{\widehat{\kappa}} Dz (z-\widehat{\kappa})^2 &= \frac{(1+\widehat{\kappa}^2)\left[1+\rm{Erf}\left(\frac{\widehat{\kappa}}{\sqrt{2}}\right)\right]}{2}+\frac{\widehat{\kappa} e^{-\frac{\widehat{\kappa}^2}{2}}}{\sqrt{2\pi}} \\
        &\, \\
        \int D\bm{u}\;D_\kappa^2(\bm{u})
        &= d\times\Bigg\{\,\frac{1}{2(1+\kappa_{\rm eff}^2)}\,
        \frac{\rm{IB}\left(\frac{1}{1+\kappa_{\rm eff}^2};\frac32,\frac{d-1}{2}\right)
        +\rm{IB}\left(\frac{\kappa_{\rm eff}^2}{1+\kappa_{\rm eff}^2};\frac32,\frac{d-1}{2}\right)}
        {\rm{B}\!\left(\frac{d-1}{2},\frac12\right)}\\[4pt]
        &\qquad\qquad+\frac{\kappa_{\rm eff}^2}{2(1+\kappa_{\rm eff}^2)}\,
        \frac{\rm{IB}\left(\frac{\kappa_{\rm eff}^2}{1+\kappa_{\rm eff}^2};\frac12,\frac{d+1}{2}\right)
        +\rm{IB}\left(\frac{1}{1+\kappa_{\rm eff}^2};\frac12,\frac{d+1}{2}\right)}
        {\rm{B}\!\left(\frac12,\frac{d-1}{2}\right)}\\[4pt]
        &\qquad\qquad-\frac{2\kappa_{\rm eff}}{(1+\kappa_{\rm eff}^2)\,\rm{B}\!\left(\frac12, \frac{d-1}{2}\right)d}
        \left[\left(\frac{\kappa_{\rm eff}^2}{1+\kappa_{\rm eff}^2}\right)^{d/2}
        -\left(\frac{1}{1+\kappa_{\rm eff}^2}\right)^{d/2}\right]\\[4pt]
        &\qquad\qquad+\frac12\left[1-\frac{\rm{IB}\left(\frac{1}{1+\kappa_{\rm eff}^2};\frac12,\frac{d-1}{2}\right)}{\rm{B}\!\left(\frac12,\frac{d-1}{2}\right)}\right]\Bigg\}.
    \end{split}
\end{equation}
where $\rm{IB}(z; a, b)=\int_0^z ds\,s^{a-1}(1-s)^{b-1}$ is the incomplete beta function and $\rm{B}(a, b)=\int_0^1 ds\,s^{a-1}(1-s)^{b-1}$ is the beta function. The VA capacity is $2d^2$ at $\kappa_{\rm eff}=0$ and remains $\mathcal{O}(d^2)$ for $\kappa_{\rm eff}=\mathcal{O}(1/\sqrt{d})$, it evaluates $2d$ at $\kappa_{\rm eff}=1$ and converges to $d^2/(d-1/2)\approx d$ for $\kappa_{\rm eff}\rightarrow\infty$.  

\section{Replica calculation for vocabulary classification}
\label{app:vocabulary_replica}
Using the decision-cell indicator $\Psi_k$ defined in Eq.~\eqref{eq:main_dict_psi}, the posterior volume is
\begin{equation}
    \mathcal Z
    =\int d\varrho(W)\prod_{\mu=1}^{P}
    \Psi_{k^\mu}(u_W^\mu;Z).
    \label{eq:app_dict_Z}
\end{equation}
Introducing the teacher as replica $a=0$, independence of the training patterns gives
\begin{equation}
    \overline{\mathcal Z^n}
    =\int\prod_{a=0}^{n}d\varrho(W^a)
    \left[
        \mathbb E_{\xi|Z}
        \sum_{k=1}^{D}\prod_{a=0}^{n}\Psi_k(u^a;Z)
    \right]^P,
    \qquad
    u^a=\frac{1}{\sqrt{Nd}}\sum_i W_i^a\xi_i,
    \label{eq:app_dict_replica}
\end{equation}
where the overbar averages the teacher and training data at fixed $Z$. With $\overline z_Z$ and $\Sigma_Z$ defined in Eq.~\eqref{eq:main_dict_mean_cov}, the large-$N$ fields are Gaussian, with conditional mean and covariance
\begin{equation}
    b^a=\frac{1}{\sqrt{Nd}}\sum_iW_i^a\overline z_Z,
    \qquad
    \mathcal C^{ab}
    =\frac{1}{Nd}\sum_iW_i^a\Sigma_ZW_i^{b\top}.
    \label{eq:app_dict_Q}
\end{equation}
where the common field is treated as already learned. Below we condition on its value $b$ and suppress this conditioning in the notation. We approximate the output-space covariance as diagonal in the basis $O$, introducing one overlap $q_m\in[0,1)$ per mode. Replica symmetry and the Nishimori identity imply that teacher-student and inter-replica overlaps are equal, and therefore the corresponding field representation for both of them is
\begin{equation}
    u^a
    =b+\sqrt{\frac{\operatorname{Tr}\Sigma_Z}{d}}\,
    O\left[Q^{1/2}h+(I-Q)^{1/2}x^a\right],
    \qquad
    Q=\operatorname{diag}(q_1,\ldots,q_{\mathrm{r}_\Sigma}),
    \label{eq:app_dict_spec_channel}
\end{equation}
where $h,x^0,\ldots,x^n$ are independent standard Gaussian vectors in $\mathbb R^{\mathrm{r}_\Sigma}$. Within this closure, the Gaussian pattern average in Eq.~\eqref{eq:app_dict_replica} depends on the weights through $\bm q$, and by changing variable we separate between volume of weights at a given $\bm q$ and energetic contribution of data and task
\[
    \begin{aligned}
        \overline{\mathcal Z^n}
        &\asymp\int d^{\mathrm{r}_\Sigma}q\,
        \mathcal V_n(\bm q)\,\mathcal A_n(\bm q;Z)^P,\\
        \mathcal A_n(\bm q;Z)
        &=\mathbb E_{h,x^0,\ldots,x^n}
        \sum_{k=1}^{D}\prod_{a=0}^{n}\Psi_k(u^a;Z),
    \end{aligned}
\]
We first evaluate the pattern factor $\mathcal A_n$, then the volume $\mathcal V_n$. For a fixed shared field $h$, define
\begin{equation}
    p_k(h,Z;\bm q)
    =\int Dx\,\Psi_k\bigl(u(h,x;\bm q);Z\bigr),
    \qquad
    Dx=(2\pi)^{-\mathrm{r}_\Sigma/2}e^{-\|x\|^2/2}\,d^{\mathrm{r}_\Sigma}x.
\end{equation}
The integrals over the independent $x^a$ factorize:
\[
    \mathcal A_n(\bm q;Z)
    =\int Dh\sum_{k=1}^{D}\prod_{a=0}^{n}
    \left[\int Dx^a\,\Psi_k\bigl(u(h,x^a;\bm q);Z\bigr)\right]
    =\mathbb E_h\sum_kp_k^{n+1}.
\]
Expanding $p_k^{n+1}=p_k+n p_k\log p_k+O(n^2)$ and using $\sum_kp_k=1$ gives
\begin{equation}
    \mathcal A_n(\bm q;Z)
    =1-nH_Z(\bm q)+O(n^2),
    \qquad
    H_Z(\bm q)=-\mathbb E_h\sum_k p_k\log p_k.
    \label{eq:app_dict_spec_GE}
\end{equation}
Consequently $G_E^{\mathrm{spec}}=\lim_{n\to0}n^{-1}\log\mathcal A_n=-H_Z$, and the replicated integral is now
\[
    \overline{\mathcal Z^n}
    \asymp\int d^{\mathrm{r}_\Sigma}q\,\mathcal V_n(\bm q)
    \exp\!\left\{-Nn\alpha H_Z(\bm q)+O(Nn^2)\right\}.
\]
Since input tokens lie in $\operatorname{span}(\mathbf Z)$, of dimension $\mathrm r_Z$, the fields only probe $\mathbf W_i^a\mathbf P_Z$, with $\mathbf P_Z$ the projector on this span. For each output mode $m$ we collect these weights, written in a basis of $\operatorname{span}(\mathbf Z)$, into $\bm w_m^a=\bigl(\bm O_m^\top\mathbf W_i^a\mathbf P_Z\bigr)_{i=1}^N\in\mathbb R^{M}$, with $M=N\mathrm r_Z$, and define their overlap matrix:
\[
    (R_m)_{ab}
    =\frac{1}{M}\bm w_m^a\cdot\bm w_m^b,
    \qquad a,b=0,\ldots,n.
\]
Its row and column $0$ refer to the teacher, and the others to posterior replicas. Spherical normalization fixes its diagonal entries to one; replica symmetry and Nishimori give the same off-diagonal overlap $q_m$. Then
\begin{equation}
    R_m=(1-q_m)I_{n+1}+q_m\bm 1\bm 1^\top,
    \qquad
    \det R_m=(1-q_m)^n(1+nq_m),
\end{equation}
where the determinant follows from the eigenvalues $1-q_m$ with multiplicity $n$ and $1+nq_m$ with multiplicity one. The spherical volume at fixed $R_m$ is, up to normalization,
\[
    \Omega_M(R_m)
    =\int\prod_{a=0}^{n}d^M w_m^a
    \prod_{0\leq a\leq b\leq n}
    \delta\!\left(\bm w_m^a\cdot\bm w_m^b-M(R_m)_{ab}\right).
\]
Within the entropy closure, the weight volume factorizes over output modes. Removing overlap-independent normalizations gives
\[
    \mathcal V_n(\bm q)
    \asymp\prod_{m=1}^{\mathrm{r}_\Sigma}
    \frac{\Omega_M(R_m)}{\Omega_M(I_{n+1})},
\]
so Eq.~\eqref{eq:app_dict_replica} has been reduced to
\[
    \overline{\mathcal Z^n}
    \asymp\int d^{\mathrm{r}_\Sigma}q\,
    \left[\prod_{m=1}^{\mathrm{r}_\Sigma}
    \frac{\Omega_M(R_m)}{\Omega_M(I_{n+1})}\right]
    \mathcal A_n(\bm q;Z)^P.
\]
Introducing a symmetric conjugate matrix $\widehat R_m$ through the overlap constraints makes the weight integrals Gaussian. Each of the $M$ coordinates contributes $(\det\widehat R_m)^{-1/2}$, yielding
\[
    \Omega_M(R_m)
    \asymp\int d\widehat R_m\,
    \exp\!\left\{\frac M2
    \left[\operatorname{Tr}(\widehat R_mR_m)
          -\log\det\widehat R_m\right]\right\}.
\]
The conjugate-matrix saddle is $\widehat R_m=R_m^{-1}$, hence
\[
    \log\frac{\Omega_M(R_m)}{\Omega_M(I_{n+1})}
    =\frac M2\log\det R_m+o(M).
\]
Substituting this result into the replicated integral, using $M=Nr_Z$ and $P=N\alpha$, gives
\[
    \overline{\mathcal Z^n}
    \asymp\int d^{\mathrm{r}_\Sigma}q\,
    \exp\!\left\{N\left[
        \frac{\mathrm{r}_Z}{2}\sum_{m=1}^{\mathrm{r}_\Sigma}\log\det R_m
        +\alpha\log\mathcal A_n(\bm q;Z)
    \right]\right\}.
\]
Finally,
\[
    \log\det R_m
    =n\log(1-q_m)+\log(1+nq_m)
    =n\left[\log(1-q_m)+q_m\right]+O(n^2),
\]
so the entropic contribution is
\begin{equation}
    G_S^{\mathrm{spec}}(\bm q)
    =\lim_{n\to0}\frac{\mathrm{r}_Z}{2n}
    \sum_{m=1}^{\mathrm{r}_\Sigma}\log\det R_m
    =\frac{\mathrm{r}_Z}{2}\sum_{m=1}^{\mathrm{r}_\Sigma}
    \left[\log(1-q_m)+q_m\right].
    \label{eq:app_dict_spec_GS}
\end{equation}
Combining this expansion with
$\log\mathcal A_n=-nH_Z+O(n^2)$ yields
\begin{equation}
    \begin{aligned}
        \overline{\mathcal Z^n}
        &\asymp\int d^{\mathrm{r}_\Sigma}q\,
        \exp\!\left\{Nn\Phi_Z^{\mathrm{spec}}(\bm q)+O(Nn^2)\right\},\\
        \Phi_Z^{\mathrm{spec}}(\bm q)
        &=\frac{\mathrm{r}_Z}{2}\sum_{m=1}^{\mathrm{r}_\Sigma}
        \left[\log(1-q_m)+q_m\right]-\alpha H_Z(\bm q).
    \end{aligned}
    \label{eq:app_dict_spec_phi}
\end{equation}
Differentiating with respect to an interior overlap gives
\[
    0=\frac{\partial\Phi_Z^{\mathrm{spec}}}{\partial q_m}
    =-\frac{\mathrm{r}_Z}{2}\frac{q_m}{1-q_m}
     -\alpha\frac{\partial H_Z}{\partial q_m},
\]
so the saddle-point equations read
\begin{equation}
    \frac{q_m}{1-q_m}
    =-\frac{2\alpha}{\mathrm{r}_Z}\frac{\partial H_Z}{\partial q_m}
    =-2r_Z\widetilde\alpha\,
    \frac{\partial H_Z}{\partial q_m},
    \qquad
    \widetilde\alpha=\frac{\alpha}{\mathrm{r}_Z^2},
    \qquad m=1,\ldots,\mathrm{r}_\Sigma.
    \label{eq:app_dict_unrestricted}
\end{equation}
The explicit entropic scale is $\mathrm{r}_Z\mathrm{r}_\Sigma$, of order $\mathrm{r}_Z^2$; the remaining dependence on vocabulary geometry is contained in $H_Z$. At the physical saddle $\bm q^\star$, the teacher token has conditional probabilities $p_k(h,Z;\bm q^\star)$. The Bayes decision and its error within the spectral closure are therefore
\begin{equation}
    \widehat k_{\mathrm{Bayes}}(h)
    =\operatorname*{arg\,max}_k p_k(h,Z;\bm q^\star),
    \qquad
    \epsilon_{\mathrm{Bayes}}(Z,\alpha)
    =\mathbb E_h\!\left[1-\max_kp_k(h,Z;\bm q^\star)\right].
    \label{eq:app_dict_spec_error}
\end{equation}


\end{document}